\documentclass{article}

\usepackage[english]{babel}
\usepackage{multicol}
\usepackage[letterpaper,top=65pt,bottom=40pt,left=1.5cm,right=1.5cm,marginparwidth=1.75cm]{geometry}

\usepackage{amsmath}
\usepackage{authblk}
\usepackage{graphicx}\usepackage{url}
\usepackage{xcolor}
\usepackage{latexsym}
\usepackage{subcaption}
\usepackage{multirow}
\usepackage{hyperref}
\usepackage{amsmath} 
\usepackage{float}
\usepackage{multicol}
\usepackage{subcaption}
\usepackage{graphicx}
\usepackage{array}
\newcolumntype{C}[1]{>{\centering\arraybackslash}p{#1}}
\usepackage[authoryear]{natbib}
\title{SegmentAnyBone: A Universal Model that Segments Any Bone at Any Location on MRI}

\author[1]{Hanxue Gu}
\author[2]{Roy Colglazier}

\author[1]{Haoyu Dong}
\author[1]{Jikai Zhang}
\author[1]{Yaqian Chen}
\author[2]{Zafer Yildiz}
\author[1]{Yuwen Chen}
\author[4]{Lin Li}
\author[1]{Jichen Yang}

\author[2]{Jay Willhite}
\author[3]{Alex M. Meyer}
\author[5]{Brian Guo}
\author[2]{Yashvi Atul Shah}
\author[6]{Emily Luo}
\author[2]{Shipra Rajput}
\author[6]{Sally Kuehn}
\author[6]{Clark Bulleit}
\author[6]{Kevin A. Wu}
\author[4]{Jisoo Lee}
\author[1,5]{Brandon Ramirez}
\author[1]{Darui Lu}

\author[3]{Jay M. Levin}
\author[1,2,4,5]{Maciej A. Mazurowski}

\affil[1]{Department of Electrical and Computer Engineering, Duke, NC, 27703, USA}
\affil[2]{Department of Radiology, Duke, NC, 27703, USA}
\affil[3]{Department of Orthopaedics, Duke, NC, 27703, USA}
\affil[4]{Department of Biostatistics \& Bioinformatics, Duke, NC, 27703, USA}
\affil[5]{Department of Computer Science, Duke, NC, 27703, USA}
\affil[6]{University School of Medicine, Duke, NC, 27703, USA}

\begin{document}
\maketitle

\begin{abstract}
Magnetic Resonance Imaging (MRI) is pivotal in radiology, offering non-invasive and high-quality insights into the human body. Precise segmentation of MRIs into different organs and tissues would be highly beneficial since it would allow for a higher level of understanding of the image content and enable important measurements, which are essential for accurate diagnosis and effective treatment planning. Specifically, segmenting bones in MRI would allow for more quantitative assessments of musculoskeletal conditions, while such assessments are largely absent in current radiological practice.
The difficulty of bone MRI segmentation is illustrated by the fact that limited algorithms are publicly available for use, and those contained in the literature typically address a specific anatomic area. 
In our study, we propose a versatile, publicly available deep-learning model for bone segmentation in MRI across multiple standard MRI locations.
The proposed model can operate in two modes: fully automated segmentation and prompt-based segmentation. Our contributions include (1) collecting and annotating a new MRI dataset across various MRI protocols, encompassing over 300 annotated volumes and 8485 annotated slices across diverse anatomic regions; (2) investigating several standard network architectures and strategies for automated segmentation; (3) introducing SegmentAnyBone, an innovative foundational model-based approach that extends Segment Anything Model (SAM); (4) comparative analysis of our algorithm and previous approaches; and (5) generalization analysis of our algorithm across different anatomical locations and MRI sequences, as well as an external dataset.  We publicly release our model at \href{https://github.com/mazurowski-lab/SegmentAnyBone}{Github Code}.
\end{abstract}

\begin{multicols}{2}
\section{Introduction}
\label{Intro}

Magnetic Resonance Imaging (MRI) is an integral modality in radiology and other medical specialties that provides high-quality images for the diagnosis of various medical conditions \citep{MRI_sig1, MRI_sig2}. MRI precisely depicts different tissues including bones, muscles, ligaments, and cartilage, with its non-invasive and radiation-free nature, which makes it an invaluable tool in medicine \citep{mri_adv1}. 
However, the interpretation of MRI is a labor-intensive task that requires significant expertise and is often subject to variability among readers, lacking a consistent, systematic quantitative approach.
Automated segmentation algorithms have the potential to alleviate the existing challenges, significantly enhance the efficiency of MRI interpretation, and thereby increase the value of this modality for both patients and physicians.
Particularly, the automated segmentation of human anatomy in MRI would provide significant value as it allows for quantitative analysis of the segmented anatomy. In the case of the skeletal system, this includes measurements of bone density, volume, and structural integrity, which are crucial for accurate diagnosis, treatment planning, and monitoring the progression of diseases like osteoporosis \citep{wehrli2006quantitative}. 


However, the availability of MRI segmentation models is far behind those for CT \citep{lenchik2019automated}, underscoring the challenges with the development of such models, including the limited availability of images with annotations as well as variability in the images themselves. In particular, while there are some algorithms for segmenting bones in selected MRI scans \cite{LUNDERVOLD2019102, https://doi.org/10.1002/jmri.26534}, there is no model that allows for bone segmentation across different, common MRI sequences.
Here, we propose a universal model that understands the complexities of human anatomy, starting with the bones. The designed model is capable of segmenting any bone in an MRI, regardless of the body location or MRI sequence.



The first problem we encounter in building this algorithm is the lack of well-annotated MRI data that covers clinically important anatomic structures in the human body. Although there are a few publicly available bone MRI datasets \citep{ambellan2019automated, van2023lumbar}, they primarily include scans from limited musculoskeletal locations and target one or few objects of interest. Although we can train an algorithm with the assembly of these MRI datasets,
most types of bones do not appear to be annotated in any available bone MRI datasets, \textit{e.g.,} ankle bones and hand bones.
Therefore, we decide to build a universal MRI bone dataset on our own. Through the effort of our annotation team, we collected and annotated 306 cases focusing on the T1-weighted sequence as the most standard and prevalent sequence, with 67 approved by professional readers. 

With the dataset curated, our next goal is to develop the best segmentation algorithm. In addition to the common selections, including Unet \cite{ronneberger2015u} and its variants \cite{siddique2021u}, we extensively explore applying foundation models in segmentation. Foundation models were first utilized in Natural Language Processing (NLP) and have quickly dominated the field \citep{devlin-etal-2019-bert}. 
These models acquire their knowledge from extensive datasets, enabling them to undertake a wide range of tasks rather than being limited to a single task domain. 
The recently proposed segmentation foundation model, Segment Anything Model (SAM), has achieved competitive or even superior zero-shot performance when compared to prior supervised methods, suggesting its potential for the development of a robust and generalizable algorithm for segmenting bones across multiple locations.


However, applying SAM directly to MRI exams can lead to unsatisfying results \citep{Mazurowski_2023}, due to a lack of domain-specific knowledge in medical imaging. To effectively utilize the SAM model in the field of medical imaging, adaptations such as transfer learning or model fine-tuning are necessary \citep{ma2023segment}. Some pioneering work \citep{wu2023medical} find it is effective to introduce Parameter Efficient FineTuning (PEFT) techniques with a few Adapter layers and keep the main model's pre-trained parameters frozen. Besides this strategy, we propose two additional novel training components: 
(1) a hybrid prompting strategy that allows SAM to generate masks either automatically or based on prompts. 
(2) a depth-attention branch that refines the image features by including the 3rd dimension information.
We also introduce an augmentation strategy that pairs corresponding non-T1 sequences based on registration. Specifically, the annotations from the T1 volumes are mapped to their non-T1 counterparts, \textit{e.g.,} T2, FLAIR.

To summarize, we have developed a universal bone segmentation algorithm, named SegmentAnyBone, for MRI exams, applicable across a wide range of body locations. We believe this algorithm can be directly applied to various musculoskeletal-related clinical downstream tasks through one single model. By utilizing a singular model for these diverse applications, we anticipate not only a streamlined integration into clinical workflows but also an enriched understanding of human body composition. The success of our development relies on two important factors: (1) the curation of a bone MRI dataset that comprehensively includes most types of bones and MRI sequences, and (2) a novel foundation model-based algorithm that achieves state-of-the-art performances and can be further improved with user interactions. 
To allow future research in this direction, we make our code and the pre-trained model publicly available at \href{https://github.com/mazurowski-lab/SegmentAnyBone}{Github Code}. 


\section{Related Work}

\subsection{Bone segmentation in Computed Tomography}

Skeletal segmentation in Computed Tomography (CT) has gathered more attention compared to Magnetic Resonance Imaging (MRI). Extensive studies have been conducted on the segmentation of bones in CT scans. These investigations have covered the segmentation of various human osseous structures, \textit{e.g.,} spines \citep{qadri2023ct} and femurs \citep{yosibash2023femurs}. 
Additionally, comprehensive models capable of segmenting various bones throughout the human whole-body CT scans have been developed. An early work \citep{sundar2022fully} aims to segment various human body bones in whole-body CT scans and TotalSegmentator \citep{wasserthal2023totalsegmentator} is later proposed to segment different bone structures across different locations in the human body.
 \cite{ji2023continual} also proposed a novel continual semantic segmentation method to segment multiple organs in CT, using the TotalSegmentator dataset.

While CT scans can offer precise information about the anatomic structures of the bones, the clinical usage of MRI prevails over CT in two major aspects. First, MRI utilizes imaging procedures without ionizing radiation \citep{lee20213d}. Patients do not need the additional radiation exposure of getting a CT for pre-operative planning. Second, MRI provides high-quality images with availability in 3D modeling and excels in evaluating soft tissues and medullary bone. Such quality is particularly relevant for conditions that affect both soft and hard tissues, such as arthritis \citep{ahmed2022comprehensive}, cartilage deformation \citep{cigdem2023artificial}, bone tumors \citep{eweje2021deep}, and a range of spinal pathology \citep{kim2020diagnostic}. Furthermore, accurately segmented masks also enhance the development of emerging hybrid imaging techniques including PET/MR and CT/MR fusion, both of which require precise registration between different modalities and accurate segmentation masks \citep{arabi2017comparison}.

\subsection{Bone Segmentation in Magnetic Resonance Imaging}


There are a few studies on bone segmentation in MRI studies. Early works in this direction adapt non deep-learning (DL) based methods. For example, \cite{zarychta2022atlas} uses atlas-based models to segment knees, and \cite{foster2018wrist} applies cellular automata for wrist segmentation. More recent works apply some traditional DL-based methods on different organs, including Unet on spines \citep{deng2024effective} and Transformers on knees \citep{li2023sdmt}. Nonetheless, these studies have not been made publicly available, and their focus on single-body locations prevents the application to various aspects \citep{lenchik2019automated}.


There also exist limited whole-body MRI segmentation algorithms. For instance, \cite{ceranka2020multi} introduces an improved multi-atlas scheme for segmenting the human skeletal system from whole-body MRI studies. \cite{lavdas2017fully} evaluates the performance of classification forests (CFs), convolutional neural networks (CNNs), and multi-atlas (MA) approach on the segmentation tasks for spine and pelvis in T2-weighted whole-body MRI studies, and finds that DL-based method outperforms other traditional approaches. Though these models can cover multiple bones or whole-body skeletons, they are built directly on a few sets of whole-body MRI, which is relatively less common as compared to location-specific MRI. Consequently, the adaptability of these models to scenarios involving specific anatomic locations remains uncertain. To the best of our knowledge, there is no existing model that can segment any body location across any MRI sequence.

Two main factors prevent the development of a whole-body bone segmentation algorithm.
The first one is the inherent difficulties associated with MRI bone segmentation. The water–fat interfaces, adjacent soft tissues like tendons, or focal air in the vicinity of the bone share the same intensity as (cortical) bone, leading to potential inaccuracies in segmentation results \citep{florkow2022magnetic}. More importantly, there is a lack of publicly available MRI datasets with annotated bone masks, which further complicates the development and validation of automated segmentation models for MRI bone imaging. While CT segmentation algorithms benefit from a plethora of publicly accessible datasets with manually annotated masks, there are very few known datasets offering this essential resource in the MRI field \citep{ambellan2019automated, van2023lumbar}. 

\subsection{Foundation models in Medical imaging}

Foundation models are Transformer-based networks pre-trained on extensive datasets and have demonstrated their outstanding performance across a broad spectrum of tasks, especially in zero-shot and few-shot scenarios \citep{cheng2023sammed2d}. These models have achieved remarkable success in various domains, including NLP with the well-known ChatGPT system, as well as in the Computer Vision field with models including Segment Anything Model (SAM) \citep{kirillov2023segment} and DALL-E \citep{ramesh2021zeroshot}. Due to their exceptional generalizability, there has been an increasing trend in adapting foundation models to medical imaging segmentation tasks recently \citep{Mazurowski_2023,wu2023medical,HUANG2024103061}. However, most of these works are prompt-based adaptations \citep{cheng2023sam} without any modification of network architecture, which requires manual prompts on each object. Furthermore, the 2D structure of the original SAM would limit its effectiveness on 3D objects. To address these problems, we propose a fine-tuned strategy that allows the foundation models to support both automatic and prompt-based segmentation, and introduce a 3D volume-level attention branch.

\section{Dataset Preparation}
\label{dataset}
\begin{figure*}
    \centering
    \includegraphics[width=2\columnwidth]{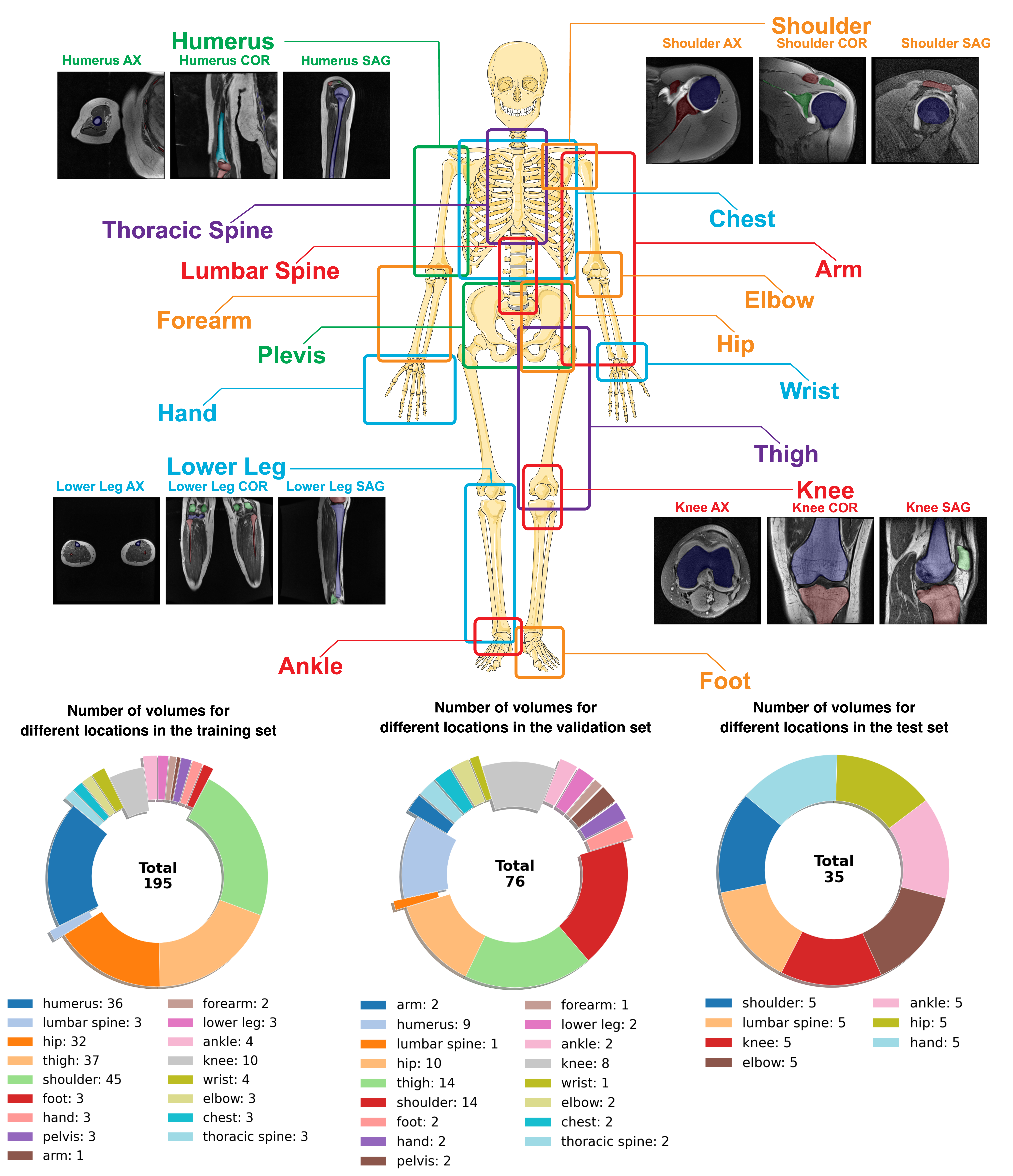}
    \caption{Dataset visualization and composition. The training dataset contains 17 locations: \textit{Humerus, Thoracic Spine, Lumbar Spine, Forearm, Pelvis, Hand, Lower Leg, Ankle, Shoulder, Chest, Arm, Elbow, Hip, Wrist, Thigh, Knee}, and \textit{Foot}. For \textit{Humerus, Shoulder, Lower Leg}, and \textit{Knee}, examples of MRI slices from three different views, Axial (AX), Coronal (COR), and Sagittal (SAG), are shown.}
    \label{fig:dataset}
\end{figure*}

In this section, we introduce the pipeline of building the datasets toward building SegmentAnyBone, including the data collection, split, annotation, and extension. We also collect an external dataset for testing.
\subsection{Data collection}

\label{sec:datacollection}
We initially employed a software query system at our institution to collect MRI radiology notes from the year 2018 and then retrieved the corresponding images. We categorized these MRI exams into 17 distinct locations within the body, as shown in Figure. \ref{fig:dataset}, using location-related keywords appearing in protocol descriptions. 
For example, if one exam has the protocol description ``MRI hip left with and without contrast'', it was assigned to the \textit{Hip} location.

For independent validation, we gathered an additional test set, comprising images spanning January 1st, 2016, to December 31st, 2020. From these, we randomly selected 5 T1 sequence volumes for each of the 7 chosen locations in the body, ensuring that there was no patient overlap with the development sets in the training phase or within the test set itself. The 7 locations were \textit{Shoulder, Elbow, Hip, Lumbar Spine, Ankle, Hand, and Knee}. They were chosen to cover the majority of the human body as well as to represent commonly acquired MRI exams. This process resulted in a total of 35 unique volumes from 35 different patients. In this test set, all annotations were approved by experienced readers.

\subsection{Dataset split for model building}
Our dataset (excluding the test set), comprising 271 annotated volumes, is divided into a training set and two distinct validation sets. Each set contains data from unique patients to ensure there is no overlap. As depicted in Figure \ref{fig:dataset}, the training set includes 195 volumes from 35 patients. The first validation set consists of 44 volumes from 17 body locations and is primarily utilized for model selection during the training phase. 
The second validation set contains 32 volumes from 5 different body locations. The annotations on these volumes have been validated and approved by experienced radiologists or physicians. This set is specifically employed as the pseudo-test set for conducting intermediate evaluations of the models.

\subsection{Image annotation}
\label{sec:annotation}
We engaged a multi-tier annotation team for image segmentation. The annotation process was guided by a fellowship-trained musculoskeletal radiologist as well as a senior orthopedic surgery fellow, experienced in cross-sectional imaging. The bulk of the annotations were performed by researchers working on the project with limited prior radiology experience who gained their expertise in bone annotation throughout the project. The experienced physicians oversaw annotation training, refinement, and approval of the final quality of many of the annotations. All test cases were approved by one of the two experienced physicians to ensure the accuracy of our evaluation.

Given the substantial volume of labeling work and the limited number of experienced annotators, we have established specific criteria for selecting MRI volumes for annotation. Firstly, we prioritized T1 sequences due to their commonality across various MRI exams and the clarity they provide for bone segmentation. Second, our initial annotation efforts concentrated on five body locations: \textit{Hip, Shoulder, Humerus, Knee}, and \textit{Thigh}. The collection of exams from these five body locations is defined as our primary dataset. Then, our annotations are expanded to additional locations with fewer volumes selected.

During the annotation process, we aimed to identify all types of bones presented in an image. 
For instance, in \textit{Shoulder} exams that incidentally include the patients' ribs, we would annotate all visible ribs accordingly. This operation allows for a consistency of model to segment any bone appearing in images across different locations.
In total, we collected 96 unique types of bones. We first aim at developing a binary segmentation algorithm of ``bone'' vs. ``not bone'' and leave the multi-class segmentation of bones as our next step. 


\subsection{Extending the dataset with non-T1 sequence}
\begin{figure}[t]
    \centering
    \begin{subfigure}[b]{0.235\textwidth} 
        \includegraphics[width=\textwidth]{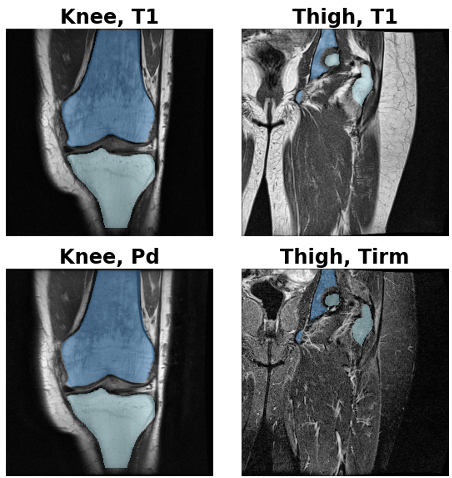}
        \caption{Well-aligned T1 and Non-T1 Pairs}
        \label{fig:well-aligned}
    \end{subfigure}
    \hfill 
    \begin{subfigure}[b]{0.235\textwidth} 
        \includegraphics[width=\textwidth]{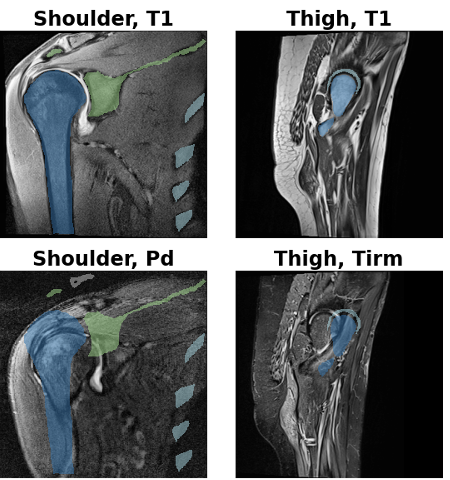}
        \caption{Poorly-aligned T1 and Non-T1 Pairs}
        \label{fig:poorly-aligned}
    \end{subfigure}
    \caption{Comparative demonstration of alignment quality. We enhanced the image contrast for better visualization.}
    \label{fig:alignment-comparison}
\end{figure}

We included a selected set of non-T1 sequences to expand the scope of our dataset. Since collecting new annotations for non-T1 volumes would require significant time and resources, we proposed to transfer existing annotations from T1 volumes to their non-T1 counterparts. 
To apply this strategy, we first selected pairs of volumes that are from the same patient and share the same view. All selected pairs were further manually viewed by a research annotator to determine if they were aligned. Figure \ref{fig:well-aligned} and \ref{fig:poorly-aligned} show pairs that are categorized as well-aligned and poorly-aligned, respectively. This process ensured that the annotations from the T1 volume could be reliably transferred to the corresponding volume of the alternate sequence only when the alignment of all skeletal structures in the pair was deemed perfect. 
The enhanced dataset includes 108 non-T1 volumes (\textit{e.g.,} T2, Turbo inversion recovery magnitude (Tirm) and Proton density (Pd) weighted) for the training set, 14 for the first validation set, 7 for the second validation set, \textit{i.e.,} the pseudo-test set, and 27 for the test dataset. The detailed distribution of the T1 and non-T1 sequences for the training, validation, and test set are shown in the Supplementary.

\subsection{Collect external test data}
We selected the SPIDER lumbar spine MRI segmentation dataset \citep{van2023lumbar}, noted as \textit{MRI-LumberSpine}, as the external test data to validate our algorithm. This dataset contains 447 exams annotating main \textit{Lumbar Spine} bones, including all visible vertebrae and the corresponding spinal structures. To eliminate the issue that some unannotated bones, such as \textit{Ribs}, might be presented in the images, only images with segmentation masks occupying at least 5 percent of the total image were selected and further cropped by the minimal rectangle containing all annotated masks. In the end, 5497 images with paired masks were used as the external test data. This dataset contains T2 and T2 SPACE volumes in addition to the T1 volumes. 

\label{method}
\begin{figure*}[ht]
    \centering
    \includegraphics[width=\textwidth]{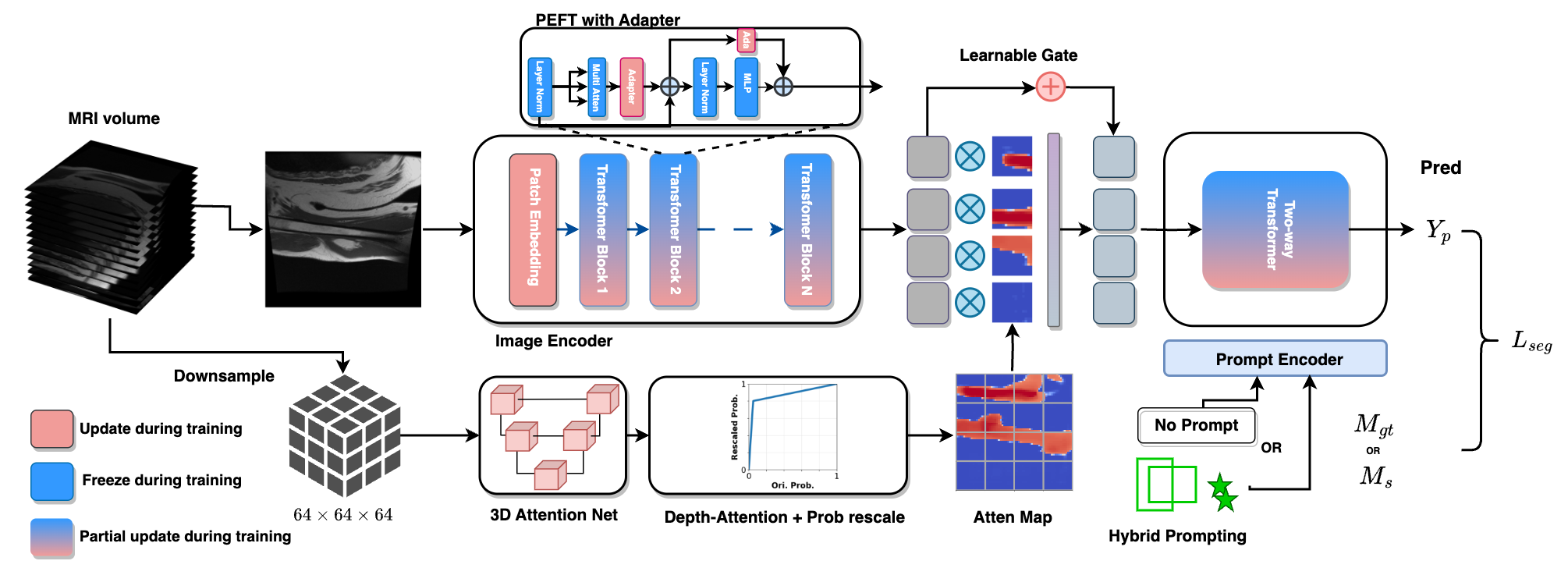}
    \caption{An overview of the model pipeline. Our model consists of the original SAM branch along with an additional 3D attention branch. These two branches are combined through a learnable gate. In the training phase, only the Adapters within the attention blocks is updated. Additionally, our approach employs a hybrid prompting technique that involves either providing specific prompts or performing automatic segmentation.}
    \label{fig:pipeline}
\end{figure*}

\section{Methods}
After curating the comprehensive dataset that contains bones across various body locations and in different sequences, our next step is to develop a versatile segmentation model capable of segmenting these bones. We first explore various well-established segmentation architectures, including CNN or Transformer-based models in both 2D and 3D segmentation settings. Next, we develop SegmentAnyBone, a novel model based on fine-tuning SAM with a newly designed prompting engineering and a depth-attention branch module that incorporates full-volume information when segmenting a single slice. 

\subsection{Standard Architectures}

\noindent\textbf{Unet}
is an encoder-decoder U-shaped structure \citep{ronneberger2015u}. It introduces the skip connection module which propagates information from the encoder layers to the corresponding decoder layers. 

\noindent\textbf{UNETR}
is a fusion of U-Net framework with Transformer models, which offers advantages in handling complex and large-scale medical imaging data \citep{hatamizadeh2022Unetr}.

\noindent\textbf{SwinUnet} 
replaces the encoder part of the Unet with the Swin Transformer that computes self-attention in an efficient shifted window partitioning scheme \citep{cao2022swin}.

\noindent\textbf{AttenUnet}
incorporates attention gates (AG) into the Unet structure \citep{oktay2018attention}. The goal of AG is to suppress irrelevant regions in an input image while highlighting salient features.

\noindent\textbf{Vnet}
increases the kernel dimension during convolution from 2 to 3 to handle volumetric (3D) medical image inputs \citep{milletari2016v}. 

\noindent\textbf{nnUnet}
can be automatically adapted to a given dataset \citep{isensee2021nnu}. It contains fixed parameters and rule-based parameters determined by the characteristics of the training dataset.
The algorithm can also make an ensemble prediction by doing cross-validation and obtaining one model from each split. We did not include this operation during implementation to make the comparison fair.

\subsection{SegmentAnyBone}
Our proposed network includes two main branches: the 2D segmentation branch built and adapted based on the Segment Anything model (SAM), and the newly introduced 3D low-resolution Attention branch.

\subsubsection{2D segmentation branch}
\noindent\textbf{SAM's Architecture.}
To elucidate our model's design and the notations used in our study, we begin with an overview of SAM's architecture. SAM consists of three primary components: (1) image encoder: a vision transformer (ViT)-based component that encodes an input image (denoted as $\mathbf{X} \in \mathbf{R}^{H\times W}$) into a latent feature embedding ($\mathbf{Z} \in \mathbf{R}^{C_Z\times H_Z\times W_Z}$). $C_Z$ represents the dimension of each embedding feature, and $H_Z$ and $W_Z$ signify the length and width of the latent feature embeddings, respectively.
In the original SAM architecture, the image encoder processes high-resolution input images ($H\!=\!W\!=\!1024$) and outputs feature maps at a scale of $1/16$ of the input image size, resulting in dimensions of $H_Z\!=\!W_Z\!=\!64$ with channel size $C=256$. (2) prompt encoder: it supports various types of prompt inputs. Depending on the type of input, it can generate sparse prompt embeddings for points or boxes or dense prompt embeddings for mask prompts. This flexibility allows SAM to adapt to different targets effectively. (3) mask decoder: it integrates embeddings from both the image and prompt encoders and outputs a multi-channel predicted mask $Y_p \in  \mathbf{R}^{C_Y\times H_Y\times W_Y} $, each channel representing a plausible prediction. 

Building upon SAM, our model's 2D-Branch retains the same input and output sizes for the image and prompt encoders as SAM, feeding each slice from the MRI volume $V$ each time and resizing it into $H\!=\!W\!=\!1024$ for network input. However, we uniquely adapt the output channels to suit our specific bone segmentation task, as the output channel is modified to cater to a binary segmentation task ($C_Y=2$) and integrate extra blocks into both the image encoder and mask decoder for fine-tuning.

\noindent\textbf{Parameter-Efficient Fine-Tuning} (PEFT) 
has been established as an effective method for refining large foundation models. This technique focuses on updating only a small fraction of the model's parameters, typically freezing over 95\% of the pre-trained parameters. This selective updating not only accelerates the fine-tuning process but also prevents the model from collapsing. 

As discussed in \citep{ma2023segment}, adding Adapter blocks to the image encoder and mask decoder works more efficiently than some other PERT techniques when fine-tuning SAM to medical imaging segmentation tasks. Thus, following a similar structure of \citep{wu2023medical}, we add Adapter blocks to SAM's attention blocks in both the image encoder and mask decoder. In SAM's image encoder, we use a binary flag in each ViT block to decide the insertion of adapters. When enabled, two distinct Adapter blocks are inserted. The first Adapter is positioned immediately after the multi-head attention head, while the second is inserted within the residual block, aligning with the multi-layer perceptron (MLP) layers. Similarly, in the mask decoder, adapters are placed after the multi-head attention blocks and within the MLP residual connections within the two-head attention. During the training process, only the parameters contained in these adapters are updated, ensuring a focused and efficient fine-tuning of the model.

\noindent\textbf{Hybrid prompting engineering.}
We use hybrid prompting engineering to add automatic segmentation to the model to compensate for the fact that SAM can only support prompt-based segmentation. During the training phase, we employ a procedure where 30\% of the batch iterations use a combination of [$\mathbf{X}$, point/box prompts, $M_s$] tuples, in line with the standard SAM fine-tuning configuration. For these iterations, we employ a dynamic prompt generation algorithm capable of creating M random point/box prompts across K disjoint regions in an image, with the condition that $M \geq K$ and K is a randomly sampled value less than the total count of separate objects present in the image, and $M_s\in M_{gt}$ is the mask covering only the selected regions and $M_{gt}$ are the ground truth mask for all targets.
Conversely, for the remaining 70\% of iterations, we omit the prompt input, effectively setting it to None, and thus only feed [$\mathbf{X}$, None, $M_{gt}$] into the model. In these automatic iterations, the prompt encoder exclusively processes the ``None'' input, resulting in the generation of default sparse and dense embeddings.

With the two types of inputs, SegmentAnyBone supports both the prompt mode and automatic mode. In the prompt mode, it retains the interactive flexibility inherited from SAM and can selectively target specific objectives among multiple objects in an image. With the addition of the automatic mode, SegmentAnyBone can autonomously segment all targeted objects efficiently. For both types, we use the same objective function $L=L_{CE}+L_{DICE}$, where $CE$ is the cross entropy loss and $DICE$ is the dice loss. When prompts appear, the target is to predict the corresponding mask regions $M_s$, and when no prompt is given, the goal is to segment all bones in a given image, \textit{i.e.,} $M_{gt}$.

\subsubsection{3D Attention branch}
\noindent\textbf{Depth-attention branch.}
During the annotation process, we observe that while radiologists primarily utilize high-resolution planes for bone analysis, they also rely on adjacent planes and need to examine how slices vary among them to ascertain the presence of bones in certain regions. This aligns with numerous prior studies emphasizing the significance of depth correlation in 3D medical image segmentation \citep{hatamizadeh2021swin, hatamizadeh2022Unetr}. 
Therefore, we propose to incorporate information from neighboring slices. Instead of following traditional 2.5D segmentation models that use multiple slices as input, or the depth-direction Adapter introduced in \citep{wu2023medical}, we introduce a novel depth-attention mechanism. This mechanism efficiently integrates depth direction information compared with the multi-slice input setting and depth-direction Adapter technique, detailed in the Supplementary.

In the depth-attention branch, for each MRI volume $V$ and its corresponding annotation $M_{V-3D}$, we initially make the volume isotropic through resizing and then downsample both the volume and mask annotation to a standardized low-resolution volume $D_r \times H_r \times W_r$ with $D_r\!=\!H_r\!=\!W_r\!=\!64$. Subsequently, we apply a lightweight 3D attention network, which incorporates a streamlined V-net architecture. This network is designed to generate over-segmented predictions in the 3D volume space since we want the predictions to cover all possible bones in a volume. 
The over-segmentation is achieved by the Tversky loss \citep{salehi2017tversky}, defined as
\begin{equation}
\begin{aligned}
    & L_{Tversky}(\alpha, \beta)  = 1 \\ 
    &- \frac{\sum_{i}^{N} p_i \cdot g_i}
    {\sum_{i}^{N} p_i \cdot g_i + \alpha \cdot \sum_{i}^{N} p_i \cdot (1 - g_i) 
    + \beta \cdot \sum_{i}^{N} (1 - p_i) \cdot g_i},
\end{aligned}
\end{equation}
where $p_i$ indicates the predicted probability over a voxel $i$, and $g_i$ indicates the corresponding ground truth. The total count of voxels is represented by $N$. Additionally, the parameters $\alpha$ and $\beta$ are used to modulate the penalties for false negatives and false positives, respectively. We set $\alpha=0.7$ and $\beta=0.3$ to prioritize a higher penalty for false negatives. Denoting the probability volume as $P_V$, the final loss function is $L_{3D}(P_V, M_{V-3D})$, where $L_{3D}=L_{CE}+L_{Tversky}$.

\noindent\textbf{Depth-attention and probability rescale.}
To compute the 2D depth-attention, denoted as $P_{attn}$, from the 3D probability volume $P_{V}$, we first remove prediction probabilities equal to or less than a threshold $\epsilon_{attn}=0.1$. These probabilities are considered indicative of regions with low interest at a high confidence level and are subsequently set to zero:
\begin{equation*}
    P_{V}(h, w, d) = 0, \ \textrm{if} \ P_{V}(h, w, d) <= \epsilon_{attn}.
\end{equation*}
This is crucial to prevent these probabilities from aggregating into significant values in the subsequent summation step. Next, the probabilities are summed along a depth range ($D_s=16$) of adjacent slices, and then normalized by the largest accumulation value.
\begin{equation*}
    P_{attn-abs}(h, w) = \sum_{d}^{D_s} P_{V}(h, w, d)
\end{equation*}
\begin{equation*}
    P_{attn}(h, w) = \frac{P_{attn-abs}(h, w)}{max(P_{attn-abs})},
\end{equation*}

where $P_{attn}(h, w) \in \mathbf{R}^{H_r \times W_r}$. As some small targets only appear in a few slices, which may lead to small attention values, we introduce a probability-rescaling function to emphasize attention for small or doubtful regions:
\begin{equation*}
    P_{attn-rescale}(h, w) = 
    \begin{cases}
        \alpha_1 * P_{attn}(h, w) & \textrm{if} \ P_{attn}(h, w) <= p_{low} \\
        \alpha_2 * P_{attn}(h, w) & \textrm{if} \ P_{attn}(h, w) > p_{low}
    \end{cases}
\end{equation*}
where $p_{low}=0.05$, and $p_{high}=0.8$, $\alpha_1 = \frac{p_{high}}{p_{low}} = 16$ and $\alpha_2 = \frac{(1-p_{high})}{(1-p_{low})}=\frac{1}{4.75}$. Similarly, to remove the noise after re-scaling, we set the probabilities not greater than $\epsilon_{rescale}$ ($\epsilon_{rescale}=1\times e^{-3}$) to 0.
\begin{equation*}
    P_{attn-rescale}(h, w) = 0, \ \textrm{if} \ P_{attn-rescale}(h, w) <= \epsilon_{rescale}.
\end{equation*}

\noindent\textbf{Attention fusion.}
Upon obtaining the depth attention, an attention gate is implemented to integrate the 3D attention into the image feature embedding, represented by $\mathbf{Z}$. The fusion process is described as follows:
\begin{equation}
Z_{fuse} = g \cdot Z + (1 - g) \cdot \mathcal{F}(Z \cdot P_{attn}),
\end{equation}
where $g$ represents a trainable parameter, initially set to 1. $\mathcal{F}$ denotes a nonlinear transformation, which is implemented using two convolutional layers with a ReLU activation function between them. The fused embeddings are fed into the mask decoder. Initially, with $g$ set to 1, the integration emphasizes the 2D branch exclusively without incorporating depth attention. This setting allows for the gradual and controlled blending of depth information into the 2D features. Additionally, it offers the flexibility to choose between a purely 2D version and a 3D-integrated version. This choice is facilitated by adjusting $g$ either as the learned weight or by fixing it at 1, thereby allowing for a customizable choice of 3D feature integration.

\begin{figure*}[t!]
    \centering
    \includegraphics[width=0.9\textwidth]{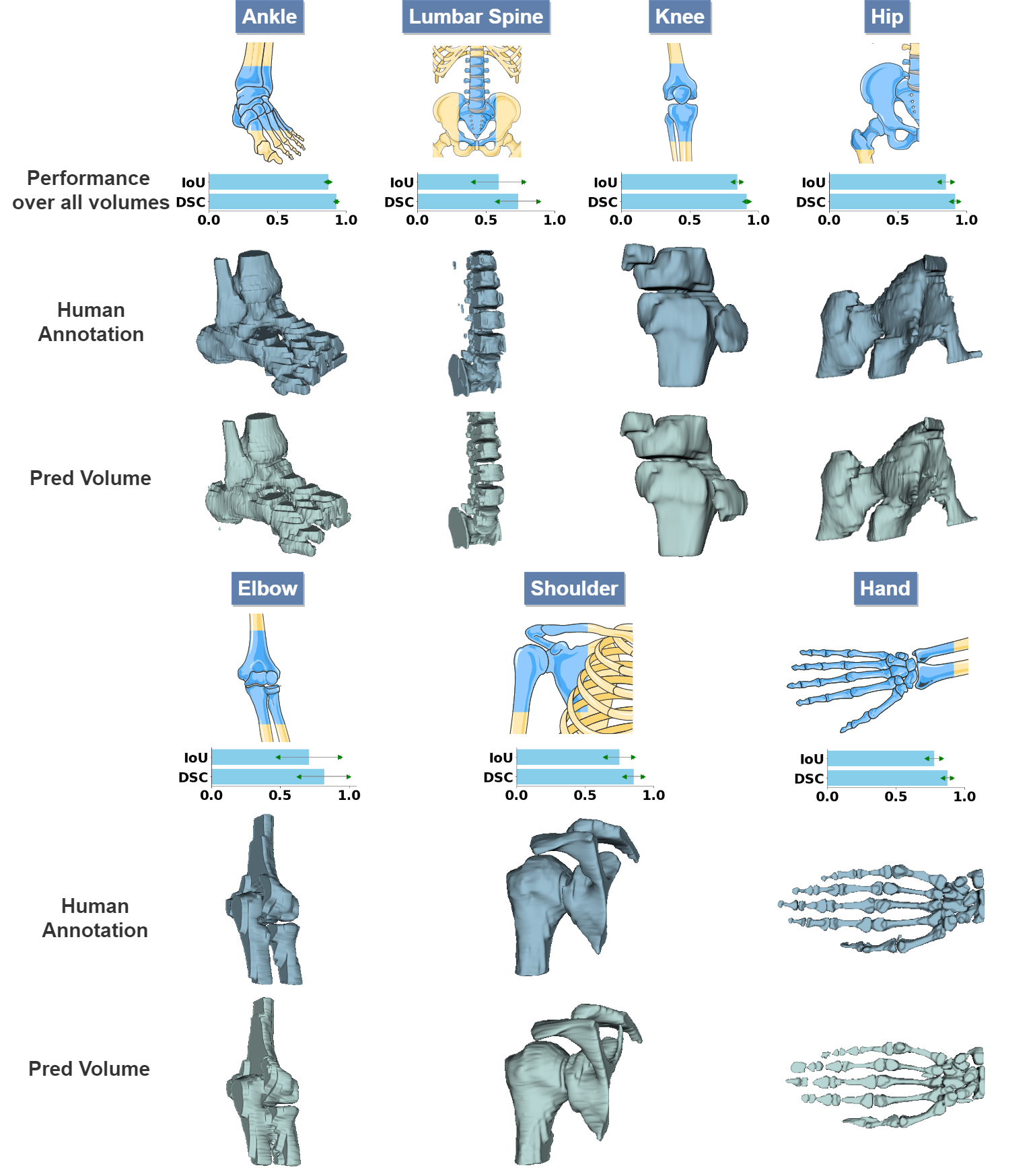}
    \caption{Visualization of the segmentation performance on the test set. The average intersection over union (IoU) and dice coefficient (DSC) scores are listed with green arrows showing the 95\% confidence interval.}
    \label{fig:vis_result}
\end{figure*}

\subsubsection{Training Pipeline.}
Throughout the training process, we selected two architectures for SAM's image encoder: ViT-base, as provided in the original paper, and MobileSAM \citep{zhang2023faster}, a lightweight ViT-based network obtained through knowledge distillation.
We initially conducted fine-tuning of the adapters in the image encoder and mask decoder in an automatic setting where training loss is $L=L(M_{gt}, Y_p)$, while the 3D attention net is separately optimized with $L_{3D}$. Following this, depth-attention maps for both the training and validation sets were pre-generated. In the second stage, the \textit{hybrid prompting} and \textit{attention fusion} were incorporated into the model. During this stage, all parameters in the image encoder, including the adapters, were frozen. The focus was solely on updating the parameters of the attention fusion Layers and the mask decoder. This bifurcated approach maintained the model's ability to make predictions without attention by ensuring that the 2D branch was thoroughly trained in the first phase. It also prevented the image encoder from becoming overly dependent on the attention mechanism, preserving its effectiveness in learning and capturing target features. In the inference stage, the model could be seamlessly toggled between a 3D attention mechanism for MRI volumes and a 2D-only mode for individual slices. This switch could be efficiently managed by adjusting the gate $g$ from a learned weight to 1, as required.

\subsection{Experimental setting}
In our experimental setup, we adapted the same configuration used in SAM for the 2D-based standard models. Specifically, inputs were resized to $3\!\times\!1024\!\times\!1024$ and their corresponding output size were maintained at $2\!\times 1024\!\times\!1024$ since there was no output downsampling. For image normalization, we applied the same mean and standard deviation values as those used in ImageNet. For the 3D-based standard models, input volumes are first resized to $128\!\times\!128 \!\times\!128$ for isotropic and then normalized to the range of [0, 1]. Regarding the nn-Unet models, which automatically adjust parameters based on the dataset, we allowed the models to autonomously determine the most suitable input and output sizes.
During all training phases, we employed various contrast and spatial augmentations to enhance model robustness. This included RandomAdjustSharpness, and RandomEqualize for contrast adjustments, alongside RandomResizedCrop and RandomRotation for spatial alterations, using standard PyTorch libraries. Additionally, we incorporated specific augmentations, such as RandGaussianNoise, RandBiasField, and RandGibbsNoise to simulate common artifacts happening for MRIs \citep{Gibson2018,morelli2011image}. Augmentations are applied in the same way for all standard architectures and SegmentAnyBone for fair comparison.

All experiments were conducted on Nvidia RTX A6000 GPUs, maintaining a batch size of 8 for a total of 200 epochs. We utilized the AdamW optimizer, with a learning rate of $5e-4$, and a warmup phase of the first 200 iterations.

\section{Experiments Results}
\begin{figure*}[ht]
    \centering
    \includegraphics[width=\textwidth]{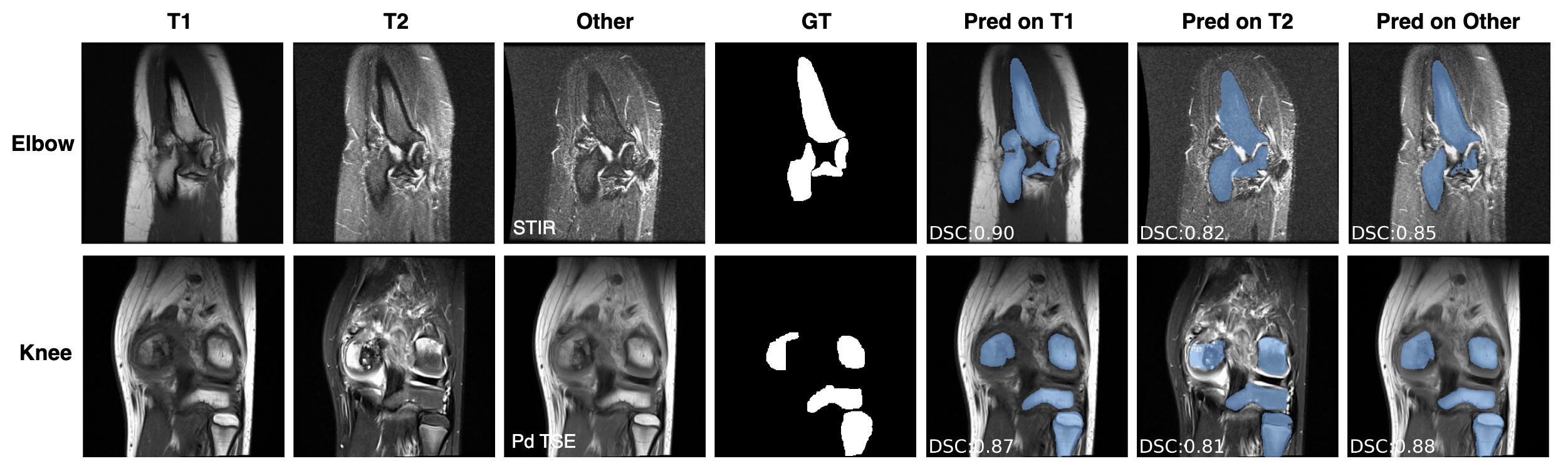}
    \caption{Visualization of the automatic segmentation performance across various sequences from the same exam. The first row shows one exam with predicted performance on T1, T2, and STIR, and the second row shows the predicted performance for T1, T2, and Proton Density TSE sequences from a \textit{Knee} exam.}
    \label{fig:multi-seq}
\end{figure*}

\begin{table*}[t]
\centering
\setlength{\tabcolsep}{2.4pt}
\begin{tabular}{l|c|c|c|c|c|c|c|c|c|c|c|c|c|c|c|c}
\hline
\multirow{3}{*}{Method} & \multicolumn{15}{c}{Body location}                                                                                     \\
\cline{2-17}
                        & \multicolumn{2}{c|}{Ankle} & \multicolumn{2}{c|}{Lumbar} & \multicolumn{2}{c|}{Knee} & \multicolumn{2}{c|}{Hip} & \multicolumn{2}{c|}{Elbow} & \multicolumn{2}{c|}{Shoulder} & \multicolumn{2}{c|}{Hand}  & \multicolumn{2}{c}{Avg}\\
\hline
 2D models                       & DSC       &  IoU         & DSC             &  IoU            & DSC         & IoU        & DSC        &  IoU        & DSC         &  IoU         & DSC           &  IoU          & DSC         &  IoU  &  DSC         &  IoU      \\
\hline
Unet                    &  73.97          &  59.03           &    52.37             &  37.39              &    50.56         &   34.88            &   63.23         &   46.28          &    55.32         &  40.53       &  55.71      &  39.00           &   80.41          & 67.64      & 61.66 &  46.39   \\
UNETR(2D)                &   68.44          &  54.08           &     32.31            &   20.48               &   48.58         &   38.96         &  64.34          &   48.51      & 47.56 & 33.96    &     48.82        &   32.81            &   76.54           &  62.32           &   55.23  &  41.59     \\
SwinUnet                &    82.82         &     70.91        &      58.13           &  43.14              &    60.94         &   49.38         &   80.84         &    68.06        &   67.60          &  53.10           &       72.65        &  57.46            &    79.96         &  66.80     &  71.85 & 58.41    \\
AttenUnet           &    82.39         &    70.30         &     57.33            &  42.14              &    60.69         &   46.48         &   73.52         &    58.30        &   64.07          &  48.33           &     59.48         &  42.60            &     78.23        &  64.44    & 67.96 &  53.23    \\
nnUnet(2D)                  &   93.34          &   87.53          &  69.81               &   56,17             &    85.07         &    77.35        &     93.43       &    87.73        &    71.84         &  59.98           &    88.94           &  80.54            &    87.20         &   77.35     & 84.23 &  74.84 \\
\hline
 3D models                       & DSC       &  IoU         & DSC             &  IoU            & DSC         & IoU        & DSC        &  IoU        & DSC         &  IoU         & DSC           &  IoU          & DSC         &  IoU  &  DSC         &  IoU      \\
\hline
Vnet  &   81.45          & 69.02           &  42.39               &  28.01              &    60.06         &   48.65         &     68.26       &   52.03         &    62.34         &    48.90         &    58.00           &   41.49            &    64.37         & 48.35       & 62.41 & 48.06  \\
UNETR(3D) &   64.90          &   48.42          &  31.98               &   19.61             &    48.48         &  32.67          &     51.13       &   34.52         &    47.10         & 31.53            &    40.38           &  25.79            &    61.50         &   44.60     & 49.35 & 33.88 \\
nnUnet(3D)                  &  88.64          & 79.62            &  59.32               &  43.80              &   70.44        &  55.14          &     80.89       &  68.07          &    69.87         &   55.78          &   76.88           &   64.35           &    67.19        & 51.15       & 73.32 & 59.70 \\
\hline

SegmentAnybone    &   93.00       &    86.88         &    73.50             &     59.14           &    91.67      &  84.66          &      91.88     &    85.00        &    81.86        &   71.10          &      85.45        &   74.87           &     87.52        &  77.90  &   86.87 &  77.08   \\

\hline
\end{tabular}
\caption{Comparison of performance across 2D and 3D segmentation methods on the test set when training on T1 and other paired sequences and testing on T1 only. ``Lumbar'' refers to ``Lumbar Spine''.}
\label{tab:comparemethods}
\end{table*}

We evaluate SegmentAnyBone in various scenarios. First, we analyze its performance across different body locations and then compare it externally with other methods. We further test its generalization ability as well as the benefits of utilizing all types of bones simultaneously over individually. Lastly, we present SegmentAnyBone's performance in the interactive segmentation mode.

\subsection{Automatic segmentation performance across different body locations}

Our first evaluation focuses on the SegmentAnyBone model, with the best-performing model trained using a combined dataset of T1 annotated cases and other paired cases.
When testing on annotated test cases on T1 sequences, the SegmentAnyBone model demonstrates strong performance on the test set, achieving an average Dice Similarity Coefficient (DSC) of 86.36\% and an average Intersection over Union (IoU) of 77.08\%. Figure \ref{fig:vis_result} presents examples illustrating our model's performance, showcasing both the quantitative evaluation using DSC and IoU metrics, and a visual representation of the predicted 3D bone volumes in comparison with human annotations. The 3D visualizations show our model's accuracy, with the predicted bone masks closely resembling the bone shapes annotated by humans, even in the complicated multi-bone and small-bone regions such as \textit{Hand} and \textit{Ankle}.

In particular, in the \textit{Ankle} area, our model achieves impressive average performance with a DSC of 93.00\% (95\% Confidence Interval (CI) [92.10\%, 93.84\%]) and an IoU of 86.87\% (95\% CI [85.66\%, 88.38\%]). The performance in the \textit{Hip} region is similarly strong, with a DSC of 91.88\% (95\% CI [88.92\%, 94.79\%]) and an IoU of 85.01\% (95\% CI [79.97\%, 90.05\%]). However, there is some variance in performance across different locations. For instance, in the \textit{Lumbar Spine} region, the model exhibits a relatively lower performance level, recording a DSC of 73.5\% (95\% CI [58.11\%, 88.66\%]) and an IoU of 59.11\% (95\% CI [40.45\%, 77.8\%]). These findings not only affirm the model's overall effectiveness but also pinpoint specific areas where further enhancements could be made.

Our model also exhibits consistent performance across different sequences. By finding the only two exams in the test set that contain two other well-aligned sequences, we show the performance of SegmentAnyBone in Figure \ref{fig:multi-seq}.
It achieves a DSC of 90.01\% for the T1 volume, 82.42\% for the corresponding T2 volume, 85.29\% on the STIR sequence for the \textit{Elbow} exam; 87.32\% for T1, 81.44\% for T2, and 88.34\% for the proton-density-weighted (PD) turbo spin-echo (TSE) sequence for the \textit{Knee} exam. Although the quality of non-T1 annotations is sub-par given they are transformed from T1 instead of being manually annotated, the high performance suggests the consistency of the performance when facing different sequences.



\subsection{Comparison with other architectures}
\begin{figure*}[ht]
    \centering
    \includegraphics[width=2\columnwidth]{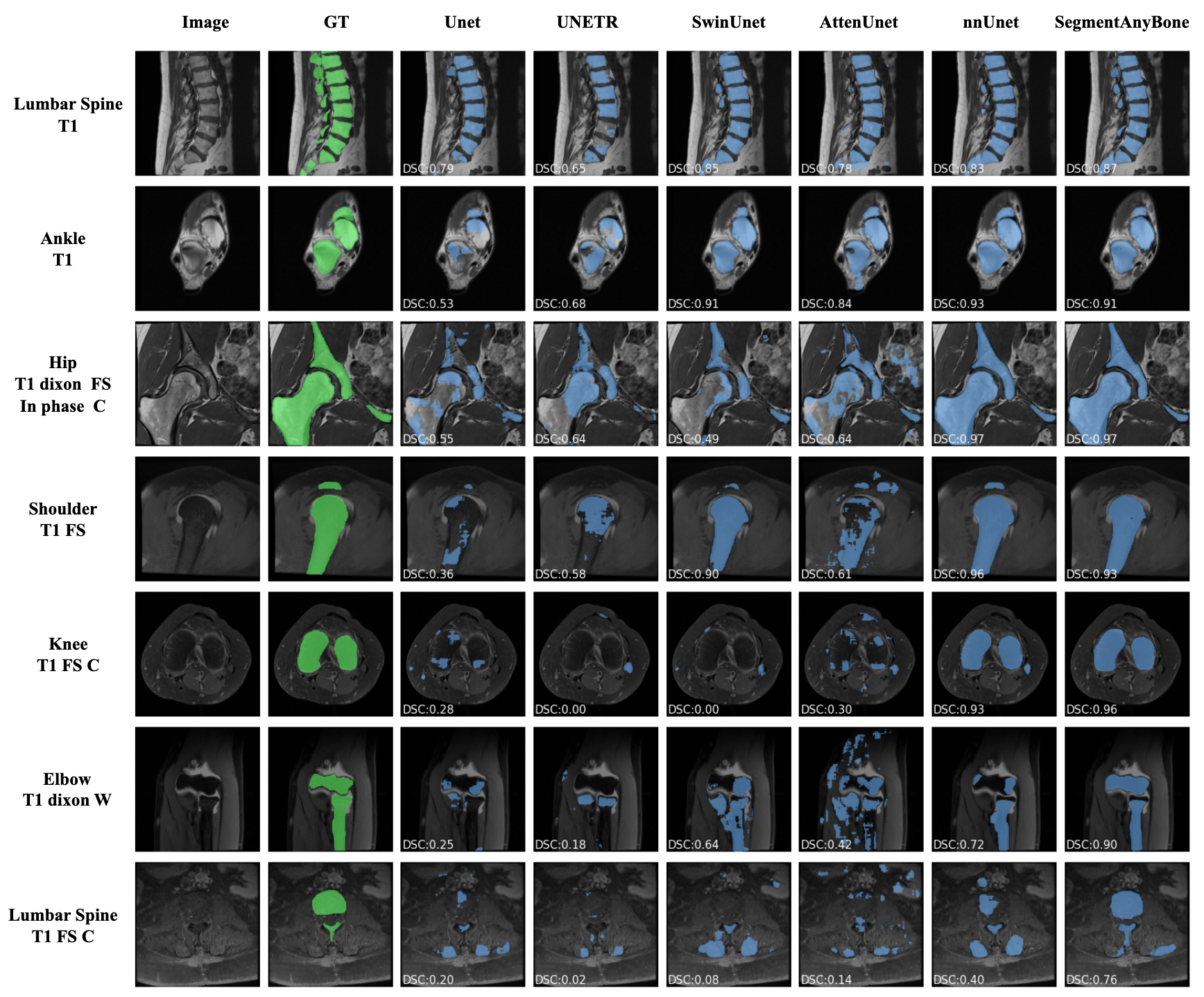}
    \caption{Visualization of the performance across 2D segmentation methods on the test set.}
    \label{fig:vis-2d}
\end{figure*}

Table \ref{tab:comparemethods} shows the performance of SegmentAnyBone with several competing algorithms in the automatic segmentation mode.
All methods share the same training set: a combination of T1 annotated cases and registered non-T1 sequences, and are tested on the same T1 test set defined in Section \ref{sec:datacollection}.
Notably, we observe that all 3D models demonstrate lower performance compared to their 2D counterparts, even when sharing a similar architecture. For instance, the nnUnet (3D) achieved a DSC of 73.32\%, notably lower than the nnUnet (2D), which reached a DSC of 84.23\%. 

When compared with 2D-based models, our model achieves a 15.02\% higher DSC and an impressive 18.67\% increase in IoU than the runner-up conventional algorithm with a fixed network structure: SwinUnet. SegmentAnyBone also surpasses nnUnet, a dynamic network that adjusts its configuration based on training data characteristics, by 2.64\% and 2.26\% in DSC and IoU, respectively. The significant improvement of nnUnet over Unet suggests the importance of finding the optimal training configuration for the target dataset, which we have not explored for SegmentAnyBone.

SegmentAnyBone particularly excels in less common sequences, such as certain \textit{Knee} and \textit{Elbow} cases that include the less common T1 sequences, \textit{e.g.,} the T1+Water sequence, achieving a 10\% improvement in DSC over nnUnet. However, in the \textit{Shoulder} region, SegmentAnyBone shows a slight decrease in precision, trailing the nnUnet by 3\% DSC. Figure \ref{fig:vis-2d} illustrates these comparisons. As shown in the Figure, in \textit{Ankle} (2nd row) and \textit{Shoulder} cases (4th row), nnUnet demonstrates greater precision in bone details. However, in the less frequent scenarios as in the \textit{Elbow} (6th row), with a darker bone appearance in T1 sequences, or in the \textit{Lumbar Spine} case (7th row) with noisy and indistinct contours, SegmentAnyBone can still make valid predictions whereas nnUnet fails.

\subsection{Evaluating the generalization ability of models}
\begin{figure*}[ht]
    \centering
    \includegraphics[width=1.8\columnwidth]{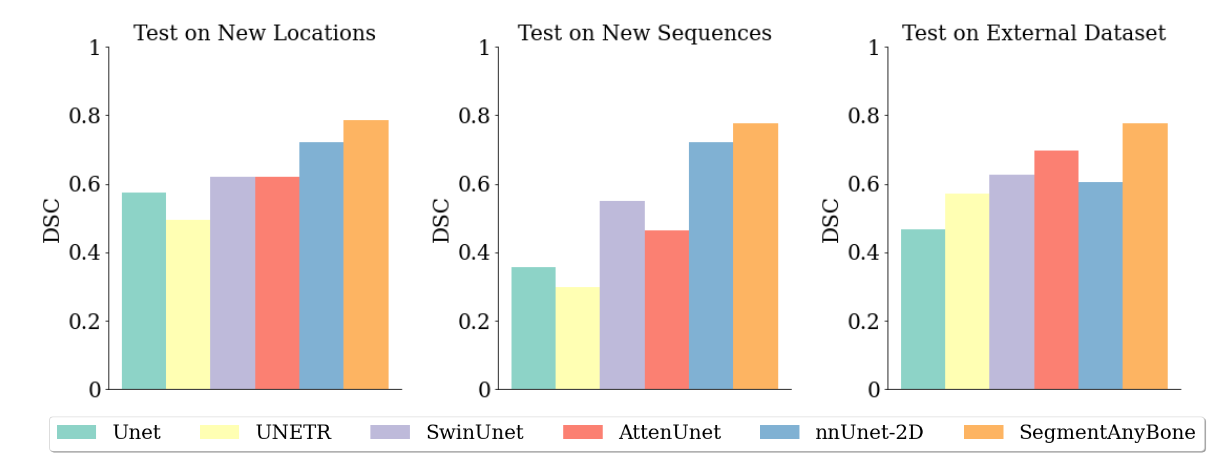}
    \caption{Analysis the generalization ability of standard methods and SegBoneMRI when testing on 1) new locations and 2) new sequences 3) new external dataset, \textit{MRI-LumbarSpine}.}
    \label{fig:generalization result}
\end{figure*}

We further assess the generalization capability of SegmentAnyBone and other algorithms through the following designed scenarios:
\begin{enumerate}
    \item Training on five primary body locations (defined in Section \ref{sec:annotation}) and tested on other locations (\textit{Ankle, Elbow, Lumbar Spine, Hand}) to assess performance on unseen targets;
    \item Training exclusively with T1 sequences and testing on different sequences to evaluate robustness to sequence variations;
    \item Training with a comprehensive set of internally sourced, annotated T1 and paired other sequences and evaluating on an external dataset \textit{MRI-LumbarSpine}.
\end{enumerate}

Figure \ref{fig:generalization result} illustrates SegmentAnyBone's performance in all scenarios when compared to other algorithms, evaluated by the dice similarity coefficient (DSC).
In the first scenario of training on five primary body locations and tested on other unseen ones, our model obtains a DSC of 78.73\%, surpassing the performances of 57.54\% for Unet, 49.34\% for UNETR, 61.89\% for SwinUnet, 61.99\% for AttentionUnet, and 71.99\% for nnUnet. 
When training on T1 sequences, SegmentAnyBone achieves a DSC of 77.68\% when facing unseen sequences of MRI data, outperforming other models with respective DSC of 35.59\% for Unet, 29.94\% for UNETR, 55.10\% for SwinUnet, 46.28\% for AttentionUnet, and 72.01\% for nnUnet. 
In tests using an external dataset from Lumbar-Spine, our model excels with the highest DSC of 77.62\%, compared to 46.64\% for Unet, 57.11\% for UNETR, 62.62\% for SwinUnet, 69.67\% for AttentionUnet, and 60.48\% for nnUnet. Overall, SegmentAnyBone, grounded in vision foundation models, demonstrates superior capabilities in handling various facets of out-of-distribution data, compared to these standard architectures.

\subsection{General model vs. specialized models}
In this section, we analyze the advantages of using exams from different locations for building one model. Besides the ``Original'' SegmentAnyBone trained with all available annotations from multiple body locations, we design two more constraint training sets:
(1) only three annotated volumes from the axial, sagittal, and coronal view from the target body location are available; and
(2) all annotated volumes from the primary five body locations (listed in Section \ref{sec:annotation}) and three additional volumes defined in (1).
Note that for these two settings, the network architecture remains the same.
We name the first setting as ``specialized'' and the second setting as ``Fine-tuned''.
For the first setting, we train the networks for 200 epochs, effectively creating a dedicated model for each location.
For the second setting, the network is initially trained with volumes from the five primary body locations and then receives a short, 10-epoch targeted adaptation. 
The short adaption period is for evaluating the model's ability to gain specific knowledge of the target location without forgetting the general knowledge from pre-training.
We repeat the experiments on four body locations: \textit{Ankle, Elbow, Lumbar Spine}, and \textit{Hand}.

The results are shown in Table \ref{tab:specificgeneral}. 
While training specialized models for individual target body locations allows for optimization in each specific area, this approach yields a comparatively lower performance, with a DSC of 78.06\%. This is less effective than ``Fine-tuned'', which achieves a 5.07\% higher DSC. 
The difference highlights the effectiveness gained from possessing prior general bone knowledge when it is applied to an unseen body area.
 ``Original'' also outperforms ``Specialized'' by 6.16\% DSC. This finding is not trivial and could eliminate our concerns about whether SegmentAnyBone, with its goal of a universal solution for all bone types, is less effective compared to a model designed for specific targets.. 

\begin{table}[]
{\scriptsize
\begin{tabular}{l|l|p{4mm}p{4mm}p{5mm}c|c}
\hline
\multirow{2}{*}{Model} & \multirow{2}{*}{Training set}             & \multicolumn{4}{c|}{Test Body Location (DSC)} & \multirow{2}{*}{AVG} \\

& & Ankle  & Elbow  & Lumbar Spine  & Hand  &                      \\
\hline
Specialized & 3 volumes                   & 92.58  & 63.26  & 70.26         & 86.16 & 78.06         \\
Fine-tuned  & Primary + 3 volumes & 92.62  & 81.78  &   71.72        &  86.50  &  83.13             \\
Original    & Full       & 93.00  & 81.86  &   74.50       &  87.52  &  84.22             \\
\hline     
\end{tabular}
}
\caption{Performance comparison with specialized model and general models. ``Primary'' refers to using the volumes from the first five annotated body locations.}
\label{tab:specificgeneral}
\end{table}

\subsection{Prompt-based Setting}
\begin{figure*}[ht]
    \centering
    \includegraphics[width=1.8\columnwidth]{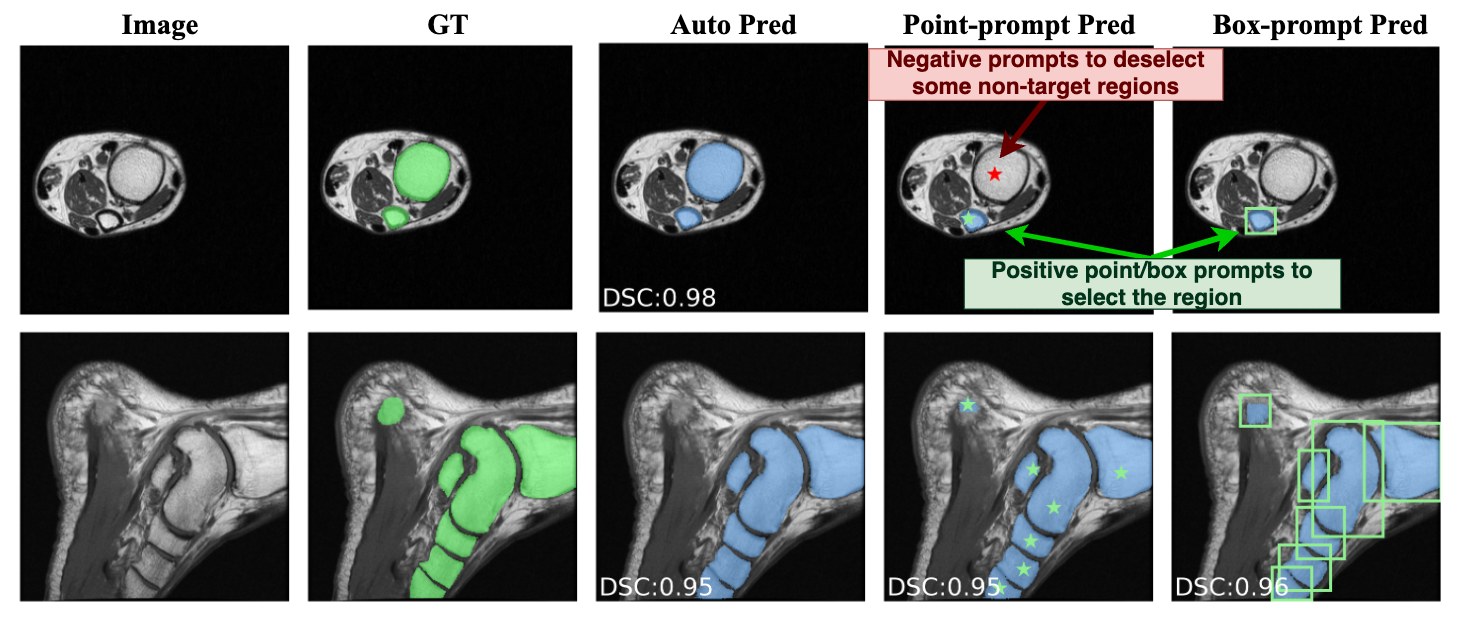}
    \caption{Visualization of result for different prompts. The first row shows the automatic segmentation of all bones, while point and box prompts target specific regions; the second-row highlights prompt-based corrections in ambiguous areas.}
    \label{fig:prompt_result}
\end{figure*}
Since SegmentAnyBone is inherited from SAM, a prompt-based segmentation algorithm, it also has the advantage of correcting the prediction based on additional prompts.
Specifically, we simulate this use case by manually putting point or box prompts on disconnected regions to get the prediction results.
For example, in automatic segmentation mode, SegmentAnyBone could segment all the bones in a slice, while putting point or box prompts allows it to only predict the desired subset of target bones, as shown in row 1 of Figure \ref{fig:prompt_result}.
Also, in the rare case when the automatic segmentation might fail to identify certain bones, we can correct the predictions through manual corrections. The minor errors in the predictions are often due to the ambiguous information in certain areas, making it challenging for the automatic process to discern the presence of bone, particularly in peripheral slices where bones begin or end. For example, as shown in row 2 of Figure \ref{fig:prompt_result}, the automatic segmentation overlooked a small fragment of the Calcaneus, and the introduction of point or box prompts in these specific areas enhanced the model's ability to correctly identify these small bone segments.

\section{Discussion and Conclusion}
We proposed SegmentAnyBone a universal model for segmenting bones across various locations in the body and various MRI sequences. To achieve this ambitious goal, we collected and annotated a new MRI dataset containing various body locations and annotated all appearing bones on these MRIs. We explored several conventional CNN-based and transformer-based segmentation models for this task, proposed a novel SAM-based algorithm using PEFT, and also introduced a new depth-attention branch. To the best of our knowledge, SegmentAnyBone is the first work that builds a universal model for segmenting tissues across all body locations.

Given the limited number of publicly available bone MRI datasets, we extensively annotated 313 volumes on 17 different main body locations on T1 sequences and manually incorporated some non-T1 exams that aligned well with the annotated ones. By training on a combination of T1 annotations and other sequences, SegmentAnyBone can achieve an average performance of 86.87\% DSC and 77.08\% IoU in segmenting bones across different body locations, a score similar to the variance caused by different annotators \citep{ozdemir2017interactive}. The 3D visualizations particularly highlight the model's precision, with the predicted bone masks closely resembling the bone shapes annotated by humans, even in the complicated multi-bone and tiny-bone regions such as \textit{Hand} and \textit{Ankle}, as shown in Figure \ref{fig:vis_result}. As shown in those 3D visual examples, though our model is built based on a foundation model utilizing slice-based prediction, it can preserve the bone's 3D structure, preventing an inconsistency between slices. We believe this property is contributed by our newly proposed depth-attention mechanism, which can integrate the 3D dimension information into a slice's feature map and share information between slices. When testing on various sequences from the same exam, our model also exhibits reliable and consistent performance across different types of sequences, as shown in Figure \ref{fig:multi-seq}. These results suggest SegmentAnyBone's capability to balance complementary information from different MRI sequences and be applied for cross-sequence analysis in complex cases where multiple imaging sequences are needed to fully understand a patient's condition. 

When compared externally, our model outperforms all other standard architectures in both 2D and 3D models when tested on the T1 sequence only. Interestingly, all 3D models demonstrated lower performance compared to their 2D counterparts, even when sharing a similar architecture. This observation can be explained by two key factors: (1) MRI exams generally exhibit lower resolution in the depth (axial) direction, resulting in less reliable information for segmentation purposes, which also demonstrates the effectiveness of our new depth-attention branch with a learnable gate; (2) 2D segmentation models benefit from a larger and more diverse training dataset, encompassing over 6000 image slices, as opposed to the 195 volumes available for 3D model training.
Among these methods, nnUnet (2D) exhibits slightly lower performance than ours. To be noticed, nnUnet had its strength by dynamically adjusting the network structure and other parameters based on the target dataset. We believe our network can be further improved with the same procedure, and leave this as a future research direction. 
Also, our model significantly excels in out-of-distribution scenarios, notably when encountering new locations, different sequences, and external datasets. This underscores the foundation model's robust zero-shot learning capabilities, establishing it as a more versatile and stable baseline for a range of downstream applications.

SegmentAnyBone also shows its effectiveness in the few-shot learning setting. When only a few volumes are annotated, a common scenario when developing an algorithm for the dataset of interest from scratch, SegmentAnyBone shows its ability to benefit from having additional information of the same type, \textit{i.e.,} the performance is boosted when the method is first pre-trained on some annotated bone volumes than from scratch. Another option is to apply our algorithm directly to a specific body location with few annotations, a feasible option since we release the code and weights publicly, as we have shown the superiority of our algorithm over its ``specialized'' version trained on volumes from a specific location only. This observation validates the benefit of achieving a global optimal model for all types of bones across different body locations over a local optimal one for a specific location.

In addition to its capability for automatic segmentation, SegmentAnyBone can also be used interactively, leveraging the prompt-based characteristics of SAM. As illustrated in Figure \ref{fig:prompt_result}, the manual input of point or box prompts can aid in correcting previous errors and achieving more accurate predictions. Beyond its primary function as a segmentation model, SegmentAnyBone shows its potential as a valuable annotation tool that can enhance the efficiency of MRI annotation.

We recognize our models' limitations in the following aspects: first, our model is configured for binary segmentation of bone/non-bone. However, we anticipate that expanding its capabilities to include bone classification should be relatively straightforward. For example, one approach could be to integrate a location-aware, smaller classifier at the end of the model's output. Second, our current testing primarily concentrates on seven body locations, a limitation dictated by the restricted availability of expert annotation time. Nonetheless, we believe that these seven locations are representative of the majority of common musculoskeletal MRI examinations, encompassing a diverse range of bone types.

In conclusion, SegmentAnyBone is a novel method that achieves state-of-the-art performance in segmenting bones on MRI across different body locations and different sequences. It can generalize well to other unseen cases and also be easily adapted to other tasks. By establishing this model and making it publicly available, we hope it can provide as a universal tool that applies to various downstream applications, reduce labor and costs in radiological measurement, and potentially uncover new insights in clinical trials and research.

\section*{Acknowledgments}
We would like to express our gratitude for the foundational Segment Anything Model, which our model is built upon, acknowledge the MONAI \citep{cardoso2022monai} framework, which provided the standard baselines crucial for our research presented in this paper, and also acknowledge Brian Lau, who helps us recruit several annotators and establish the research team for Shoulder areas. 


\bibliographystyle{apalike} 
\bibliography{refs}

\newpage
\section*{Supplementary Material}
In the Supplementary, we provide additional material for MRI volumes and patients' information to enhance the understanding of the main text and to present additional evidence to support our methods and conclusions. Section \ref{sec:dataprotocol} provides the count for different dataset protocols and different views for different body locations. Section \ref{sec:gender} demonstrates the data distribution for different patients' genders.
Section \ref{sec:ablation} presents the ablation study conducted during model development.

\subsection{Dataset Protocol and View Composition}
\label{sec:dataprotocol}
\begin{figure*}[ht]
    \centering

    \begin{subfigure}{0.99\textwidth}
        \includegraphics[width=\linewidth]{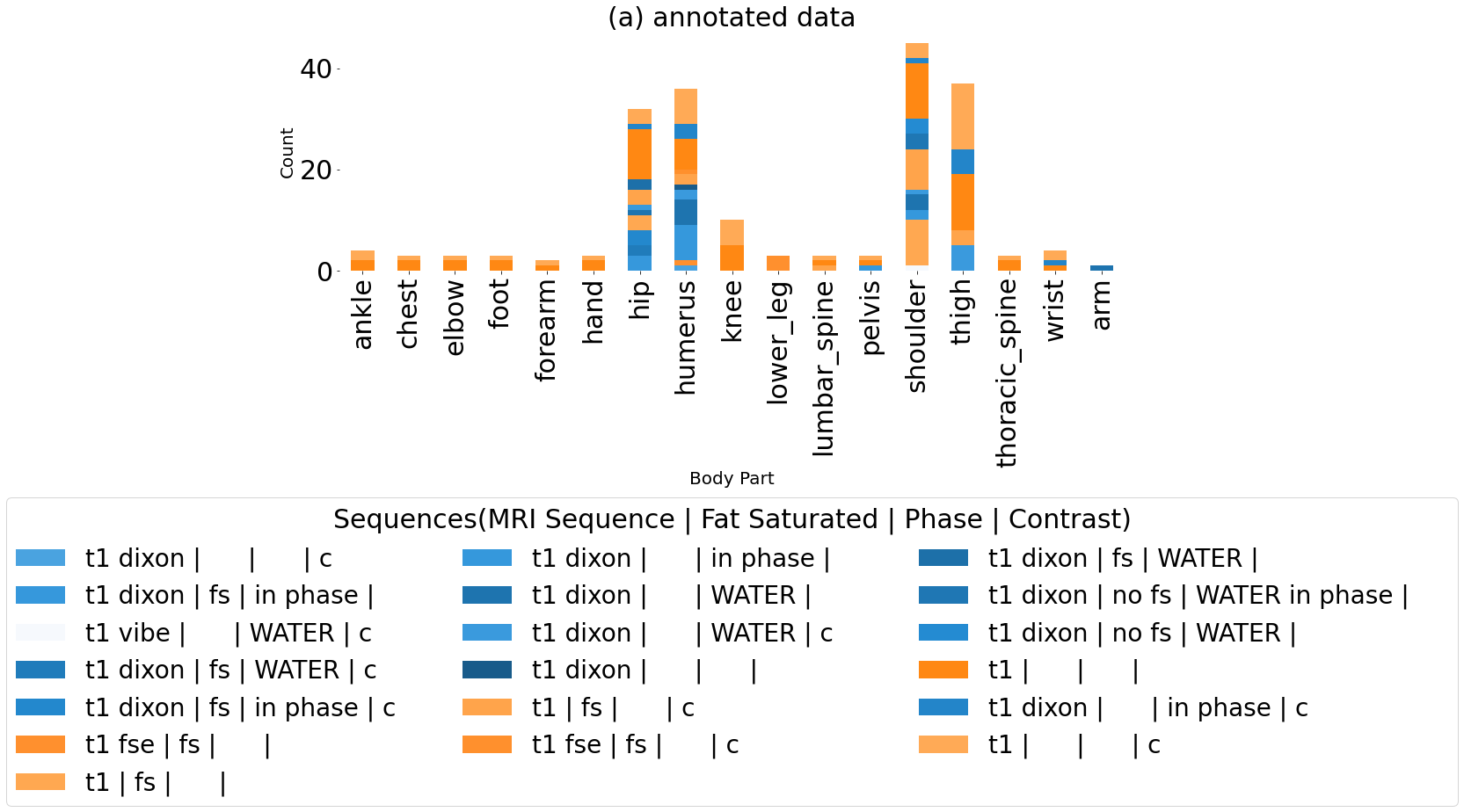}
    \end{subfigure}
    
    \begin{subfigure}{0.99\textwidth}
        \includegraphics[width=\linewidth]{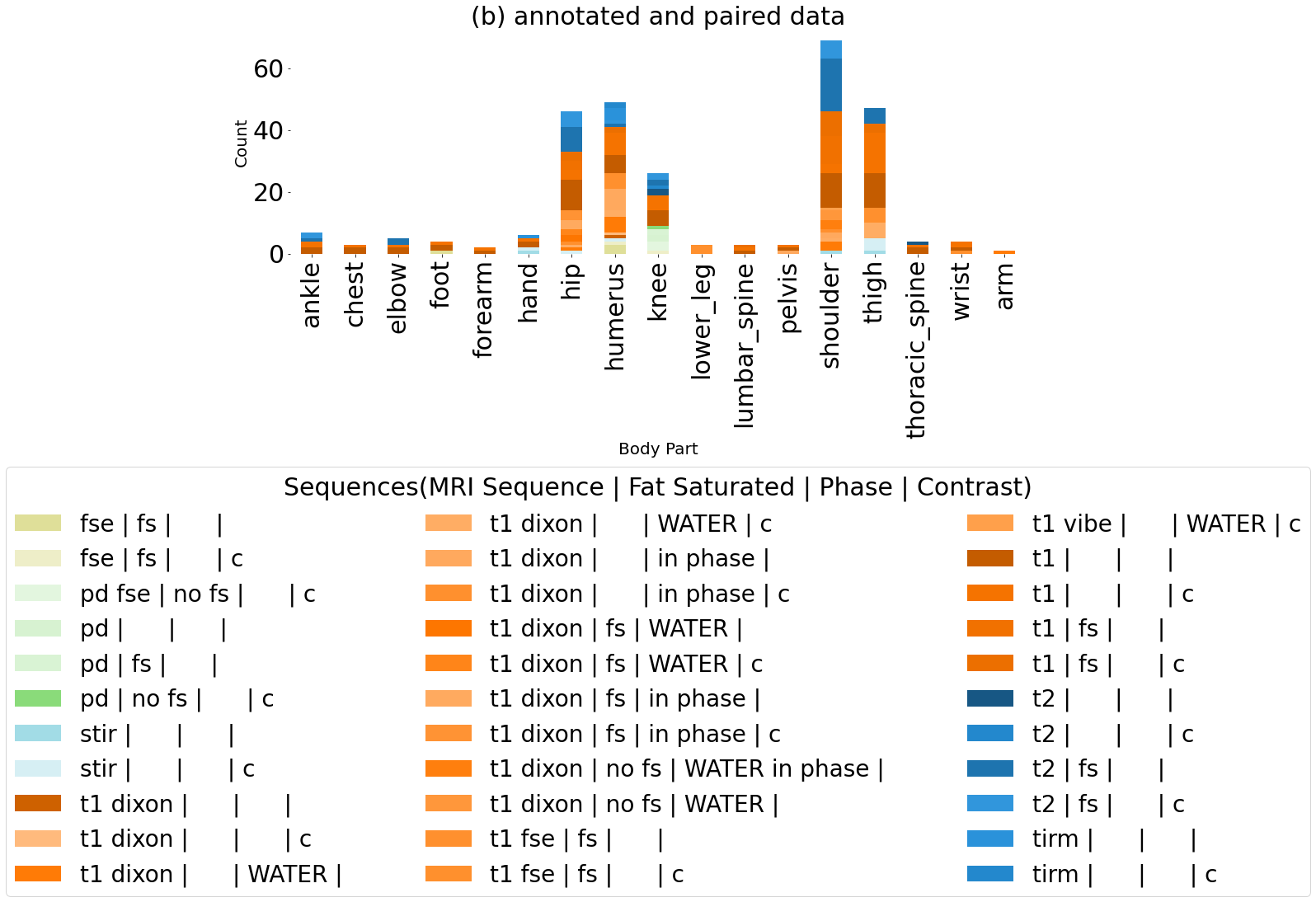}
    \end{subfigure}

    \caption{A histogram showing the distribution of sequences for different locations in the training set for (a) annotated data and (b) annotated and paired data}
    \label{fig:seq_train}
\end{figure*}

\begin{figure*}[ht]
    \centering

    \begin{subfigure}{0.99\textwidth}
        \includegraphics[width=\linewidth]{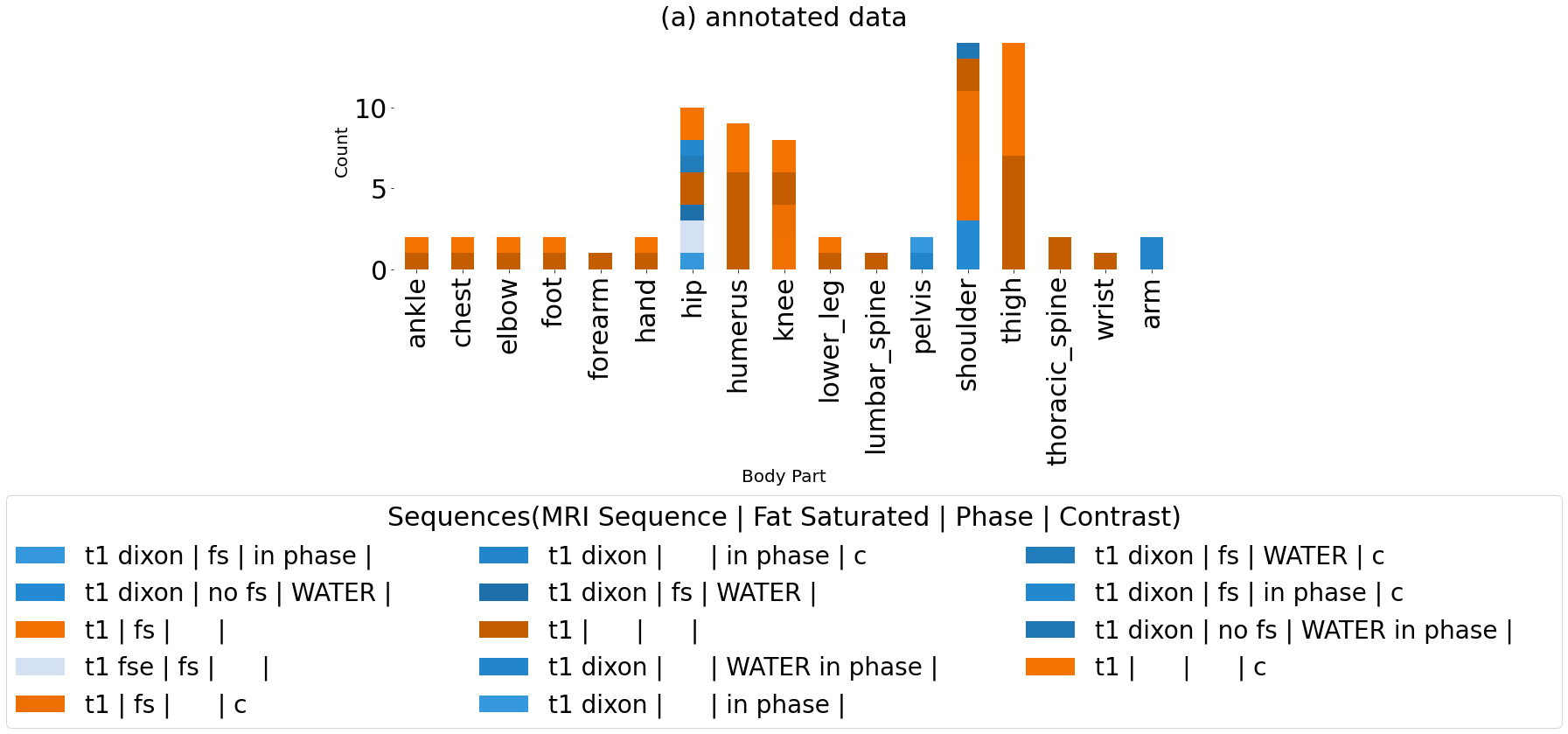}
    \end{subfigure}
    
    \begin{subfigure}{0.99\textwidth}
        \includegraphics[width=\linewidth]{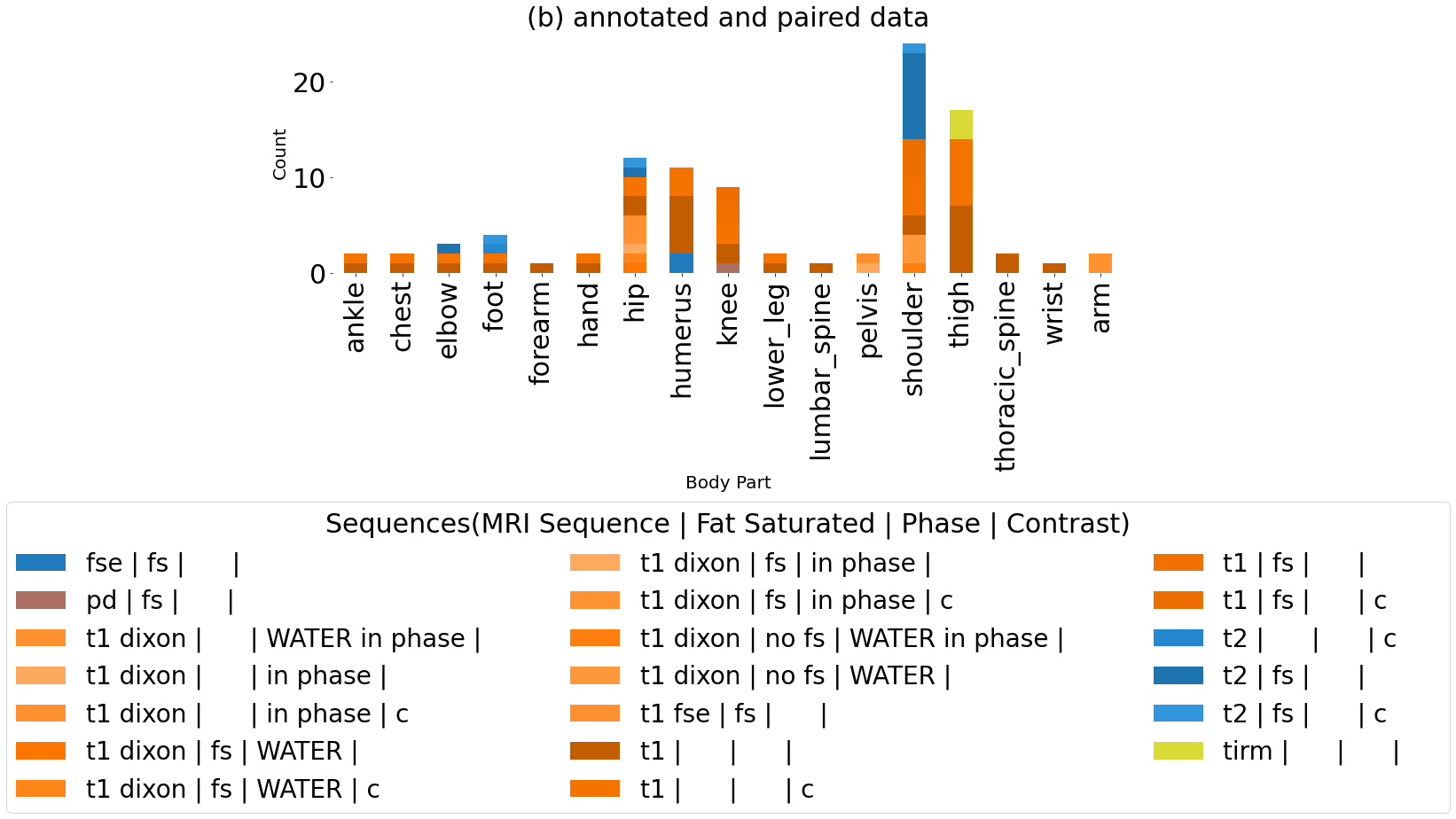}
    \end{subfigure}

    \caption{A histogram showing the distribution of sequences for different locations in the validation set for (a) annotated data and (b) annotated and paired data}
    \label{fig:seq_val}
\end{figure*}

\begin{figure*}[ht]
    \centering

    \begin{subfigure}{0.99\textwidth}
        \includegraphics[width=\linewidth]{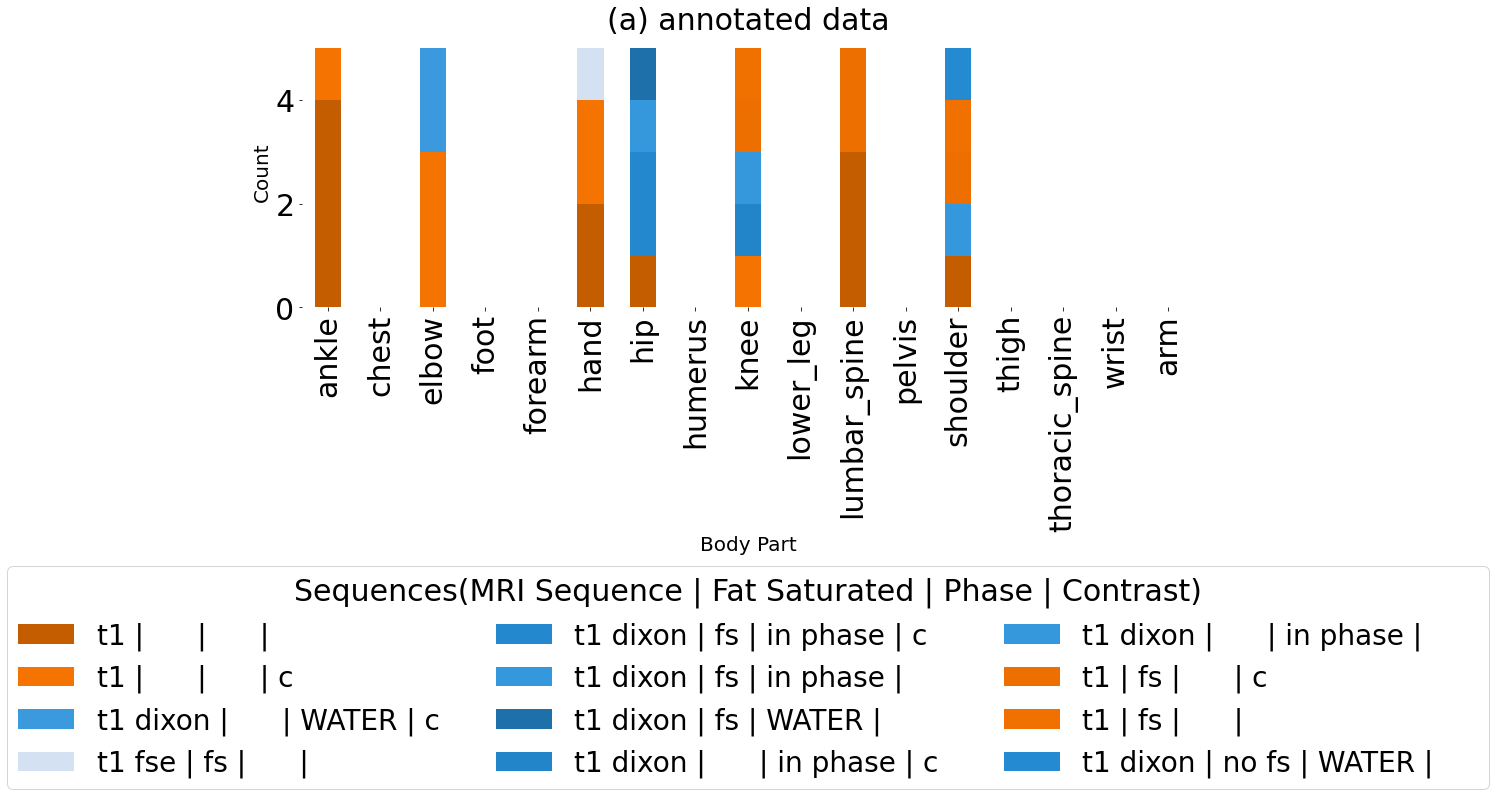}
    \end{subfigure}
    
    \begin{subfigure}{0.99\textwidth}
        \includegraphics[width=\linewidth]{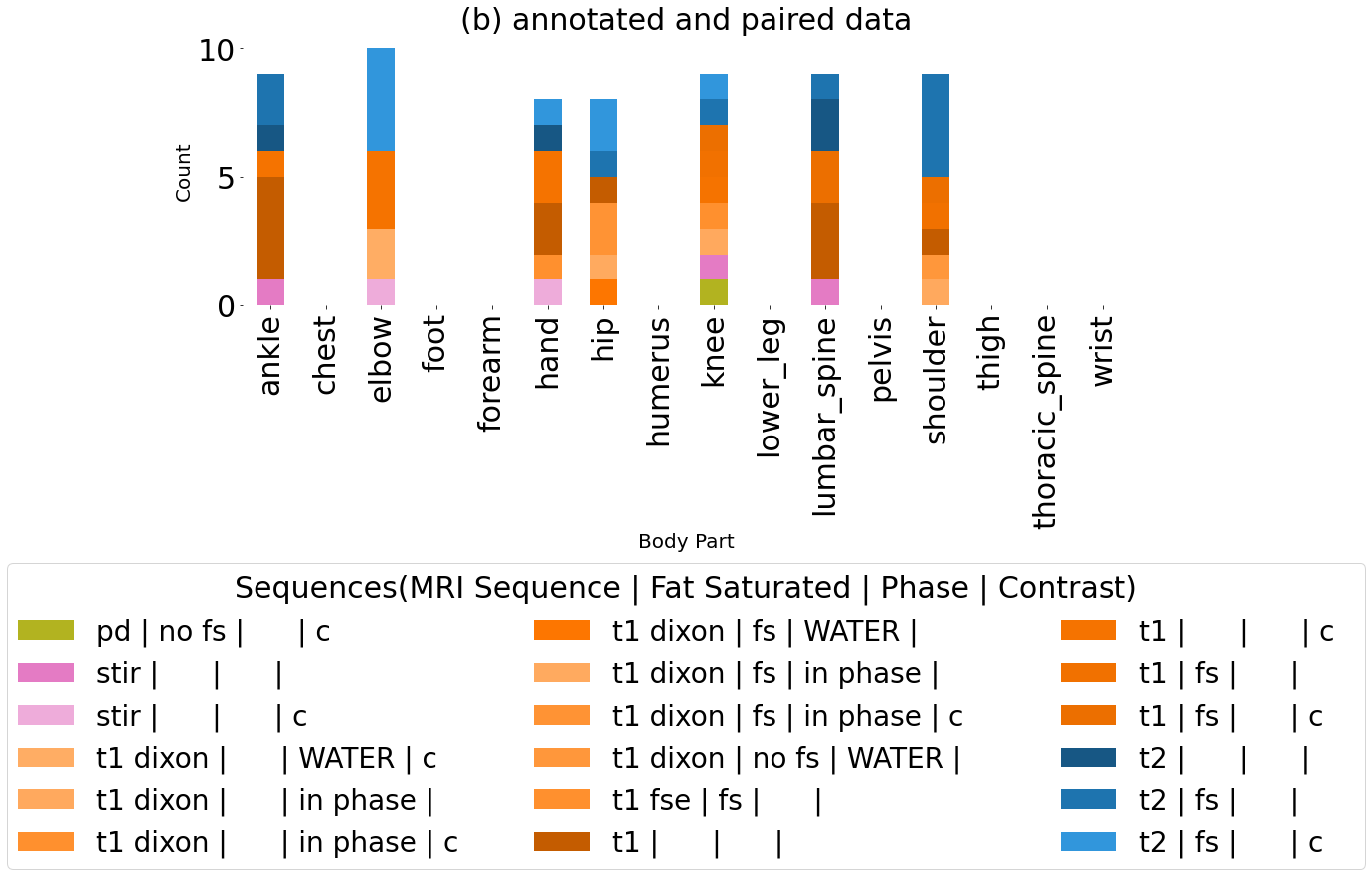}
    \end{subfigure}

    \caption{A histogram showing the distribution of sequences for different locations in the test set for (a) annotated data and (b) annotated and paired data}
    \label{fig:seq_test}
\end{figure*}

Figure \ref{fig:seq_train}, \ref{fig:seq_val}, \ref{fig:seq_test} demonstrate the sequence composition for our dataset, including annotated T1-weighted cases and manually paired non-T1 cases. Each volume has been described by a self-defined MRI protocol consisting of MRI sequence information, fat-saturated information, phase information, and contrast information. Each piece of information has been separated by `` $\mid$ ''. MRI sequence information consists ``t1'' (T1 Weighted), ``t2'' (T2 Weighted), ``pd'' (Proton Density), ``stir'' (Short Tau Inversion Recovery), ``dixon'', ``tirm'' (Turbo Inversion Recovery Magnitude), ``fse'' (Fast Spin Echo), ``vibe'' (Volumetric Interpolated Breath-hold Examination), ``blade'', ``lava'' (Liver Acquisition with Volume Acceleration), ``merge'' (Multiple Echo Recombined Gradient Echo), ``tse'' (Turbo Spin Echo). Fat saturated has ``fs'' for fat-saturated cases and ``no fs'' for non-fat saturated. Phase information describes the water phase and in phase noted as ``WATER'' and ``in phase''. Lastly, ``c'' in the description represents ``Postcontrast''. A space is used to indicate the corresponding information is not available. The colors are mainly used to indicate different types of sequences, \textit{i.e.,} ``t1'' vs. ``t2'', as to different sub-types of the same sequences.

\begin{figure*}[ht]
    \centering

    \begin{subfigure}{0.8\textwidth}
        \includegraphics[width=\linewidth]{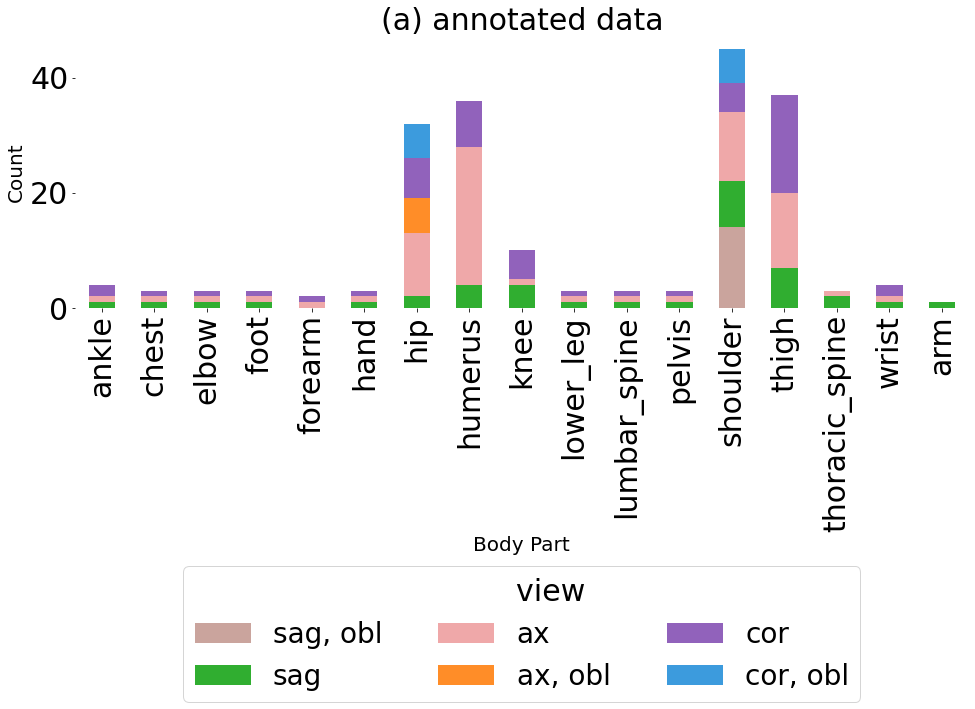}
    \end{subfigure}
    
    \begin{subfigure}{0.8\textwidth}
        \includegraphics[width=\linewidth]{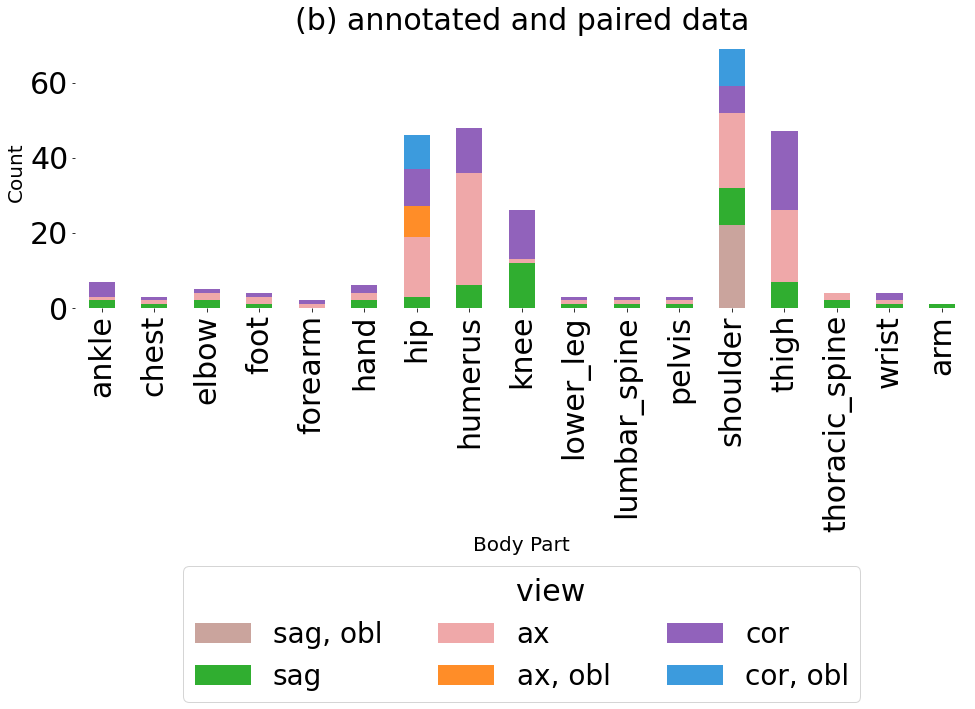}
    \end{subfigure}

    \caption{A histogram showing the distribution of views for different locations in the training set for (a) annotated data and (b) annotated and paired data}
    \label{fig:view_train}
\end{figure*}

\begin{figure*}[ht]
    \centering

    \begin{subfigure}{0.8\textwidth}
        \includegraphics[width=\linewidth]{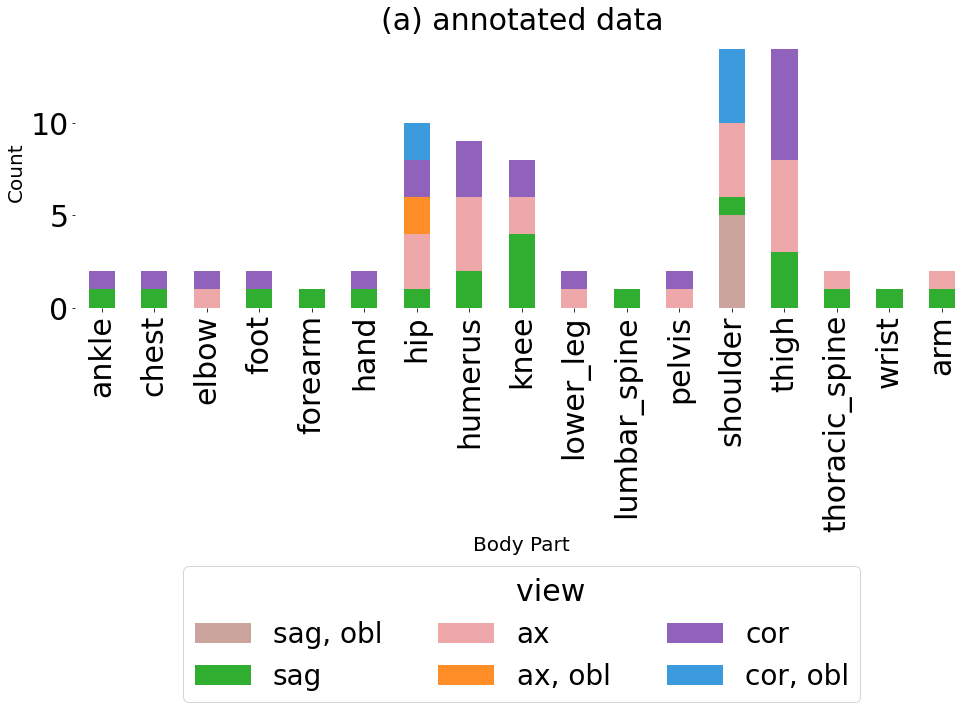}
    \end{subfigure}
    
    \begin{subfigure}{0.8\textwidth}
        \includegraphics[width=\linewidth]{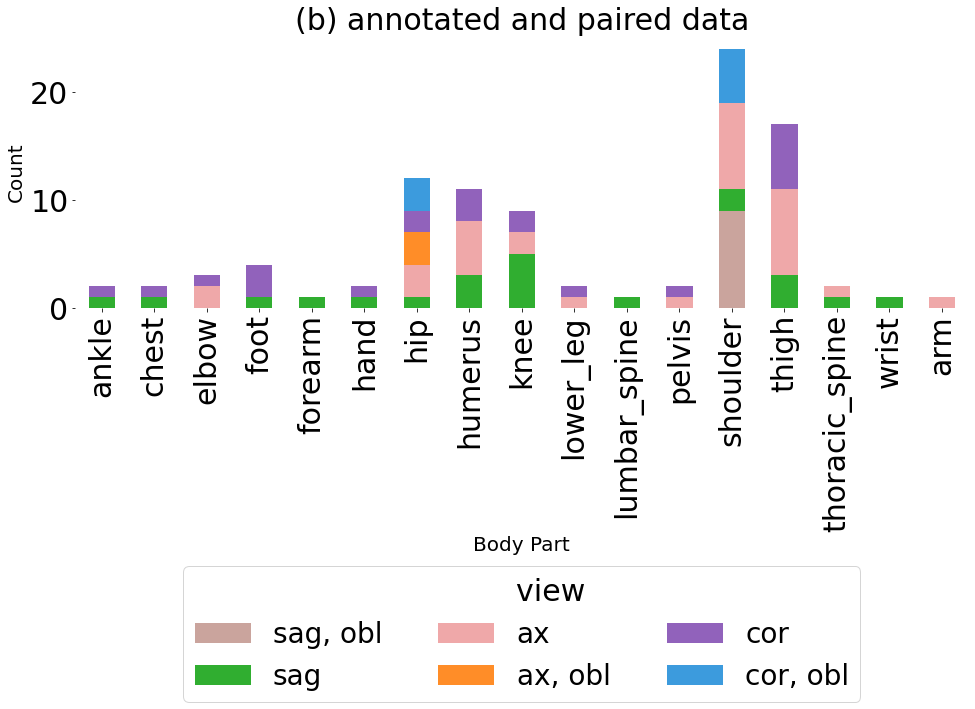}
    \end{subfigure}

    \caption{A histogram showing the distribution of views for different locations in the validation set for (a) annotated data and (b) annotated and paired data}
    \label{fig:view_val}
\end{figure*}

\begin{figure*}[ht]
    \centering

    \begin{subfigure}{0.8\textwidth}
        \includegraphics[width=\linewidth]{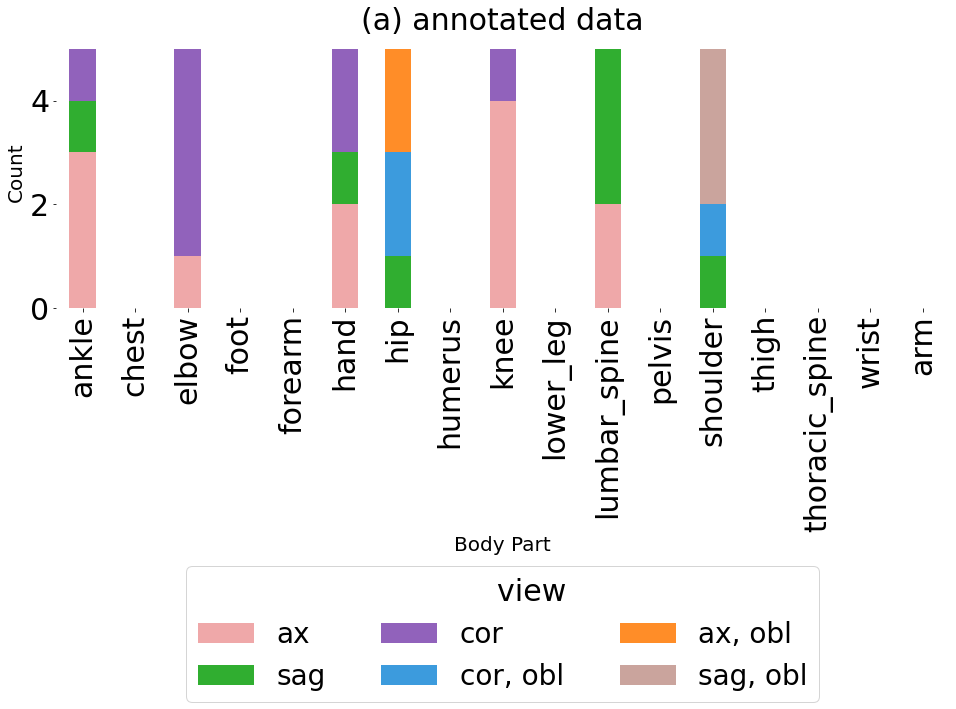}
    \end{subfigure}
    
    \begin{subfigure}{0.8\textwidth}
        \includegraphics[width=\linewidth]{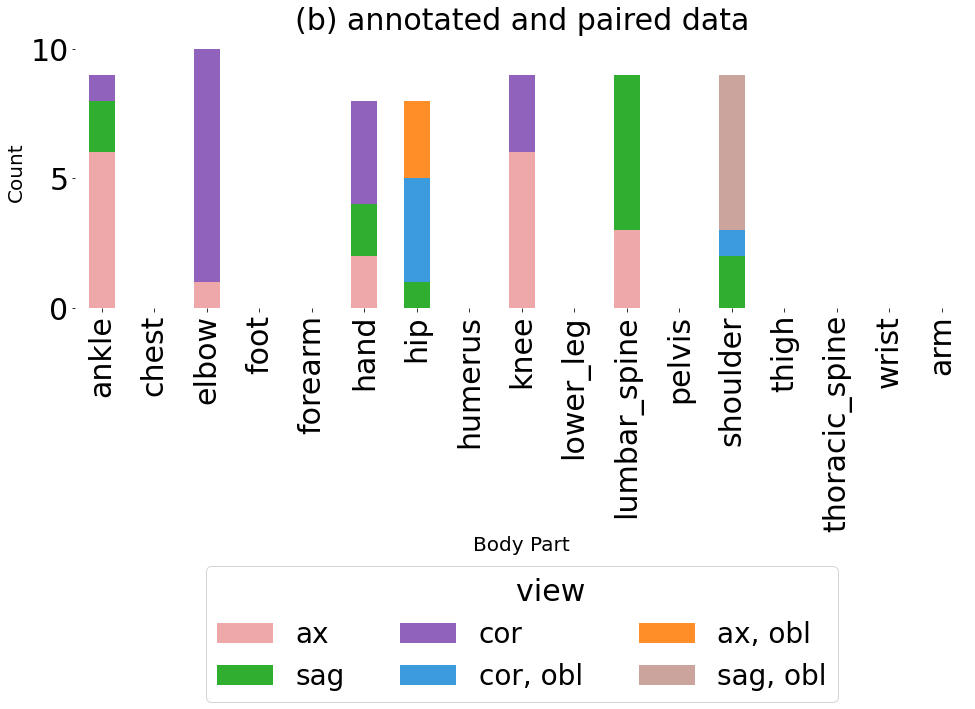}
    \end{subfigure}

    \caption{A histogram showing the distribution of views for different locations in the test set for (a) annotated data and (b) annotated and paired data}
    \label{fig:view_test}
\end{figure*}

Figure \ref{fig:view_train}, \ref{fig:view_val}, \ref{fig:view_test} also shows the view for different body locations to provide insights into our data diversity and the model's capacity for generalization. In the view description, AX is short for Axial, COR represents Coronal, and SAG describes Sagittal.

\subsection{Patients' Gender}
\label{sec:gender}
\begin{figure*}[ht]
    \centering

    \begin{subfigure}{0.4\textwidth}
        \includegraphics[width=\linewidth]{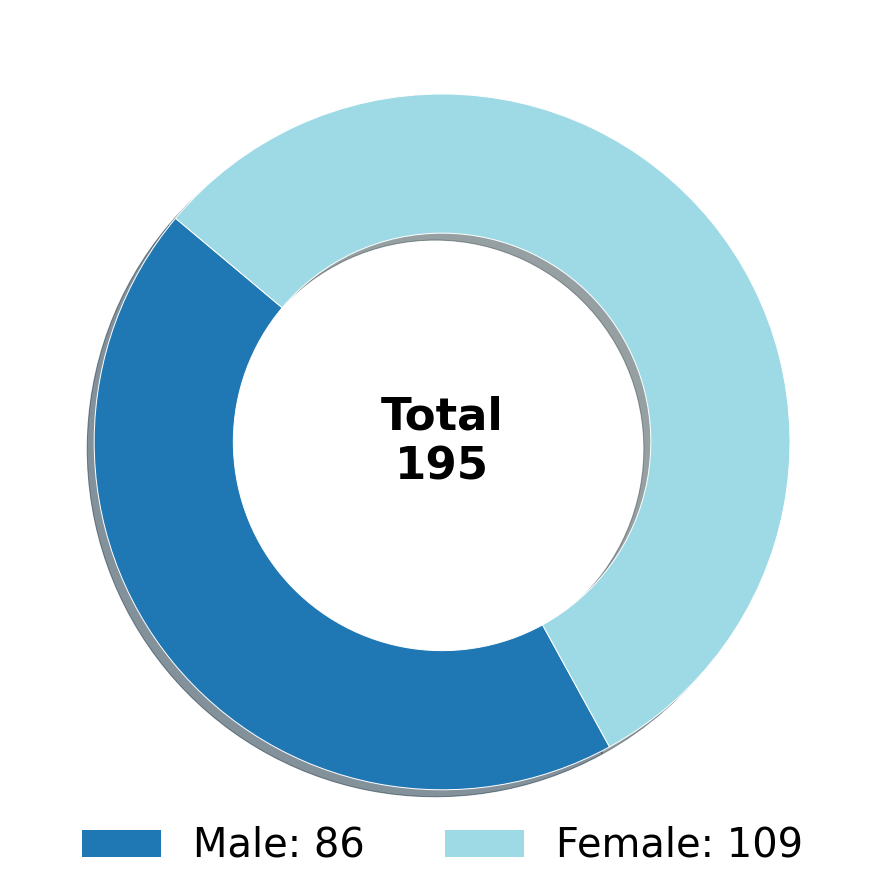}
        \caption{Count for different gender in the training set for annotated data}
        \label{fig:sub1}
    \end{subfigure}
    \hfill
    \begin{subfigure}{0.4\textwidth}
        \includegraphics[width=\linewidth]{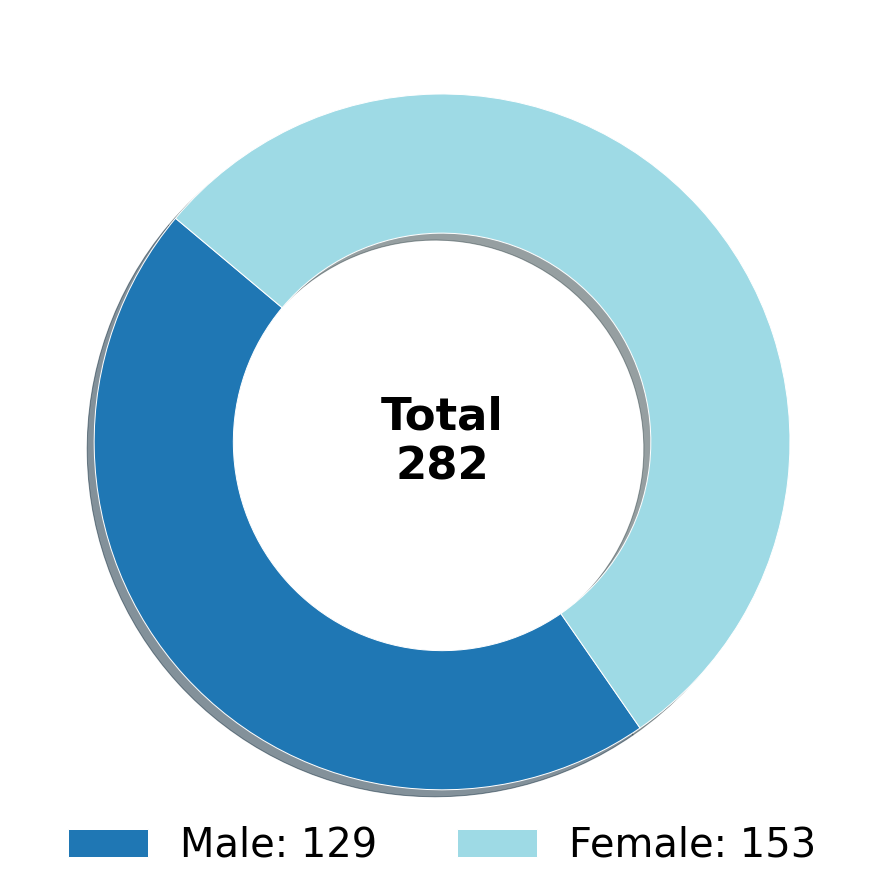}
        \caption{Count for different gender in the training set for annotated and paired data}
        \label{fig:sub2}
    \end{subfigure}
    
    \begin{subfigure}{0.4\textwidth}
        \includegraphics[width=\linewidth]{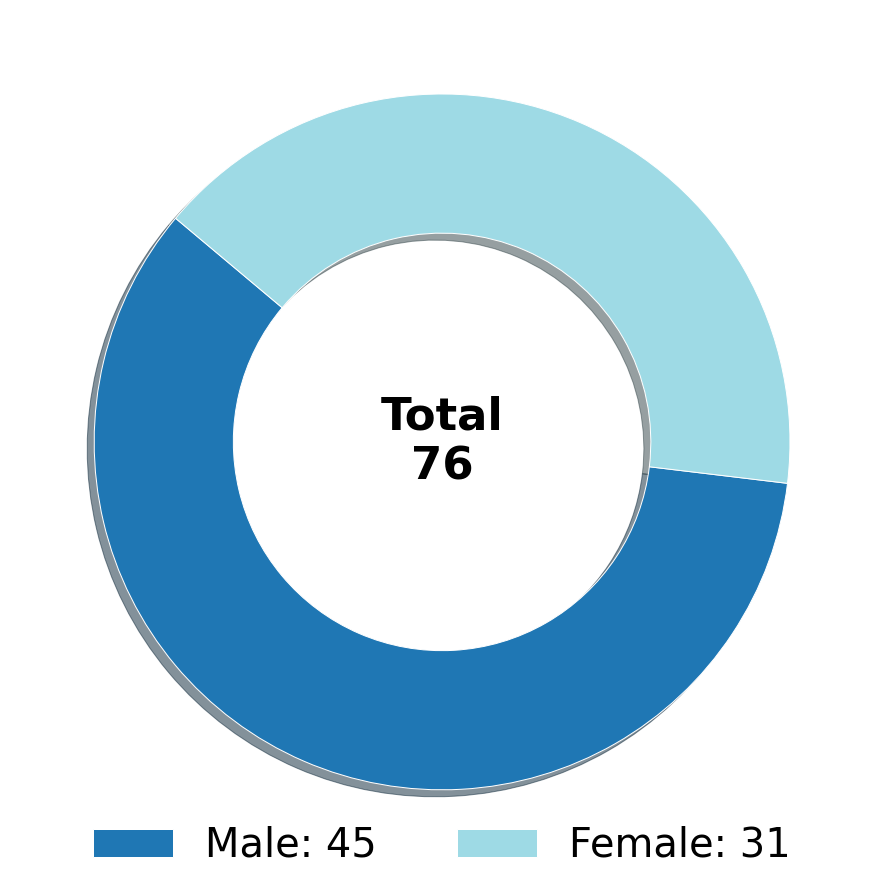}
        \caption{Count for different gender in the validation set for annotated data}
        \label{fig:sub3}
    \end{subfigure}
    \hfill
    \begin{subfigure}{0.4\textwidth}
        \includegraphics[width=\linewidth]{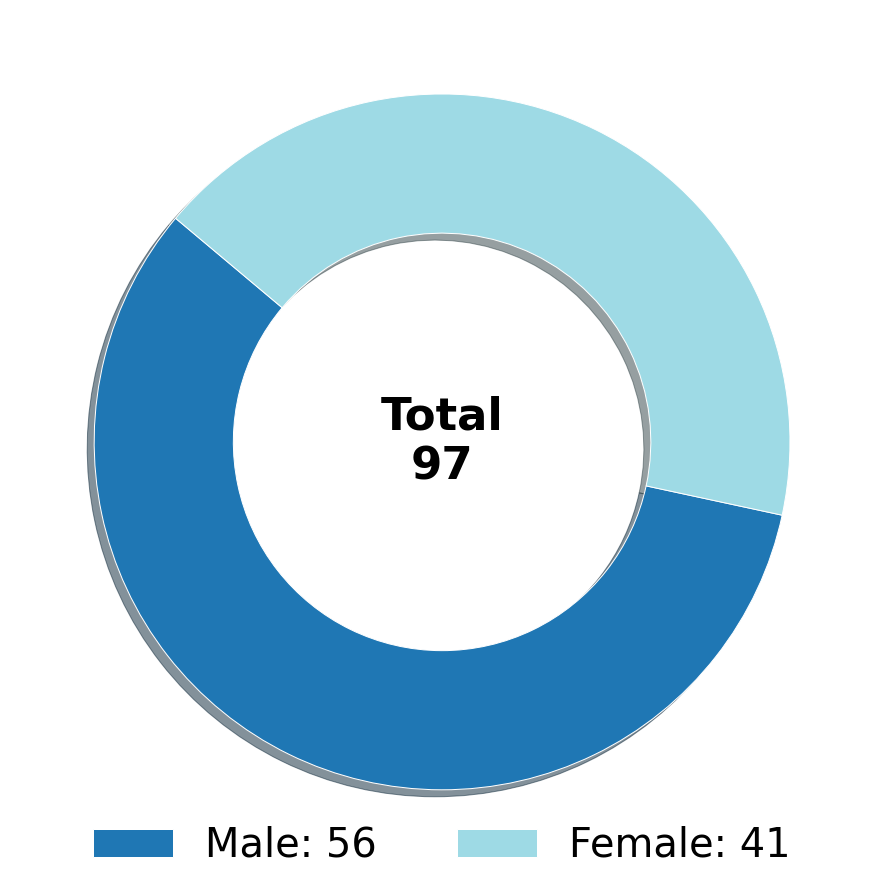}
        \caption{Count for different gender in the validation set for annotated and paired data}
        \label{fig:sub4}
    \end{subfigure}

    \begin{subfigure}{0.4\textwidth}
        \includegraphics[width=\linewidth]{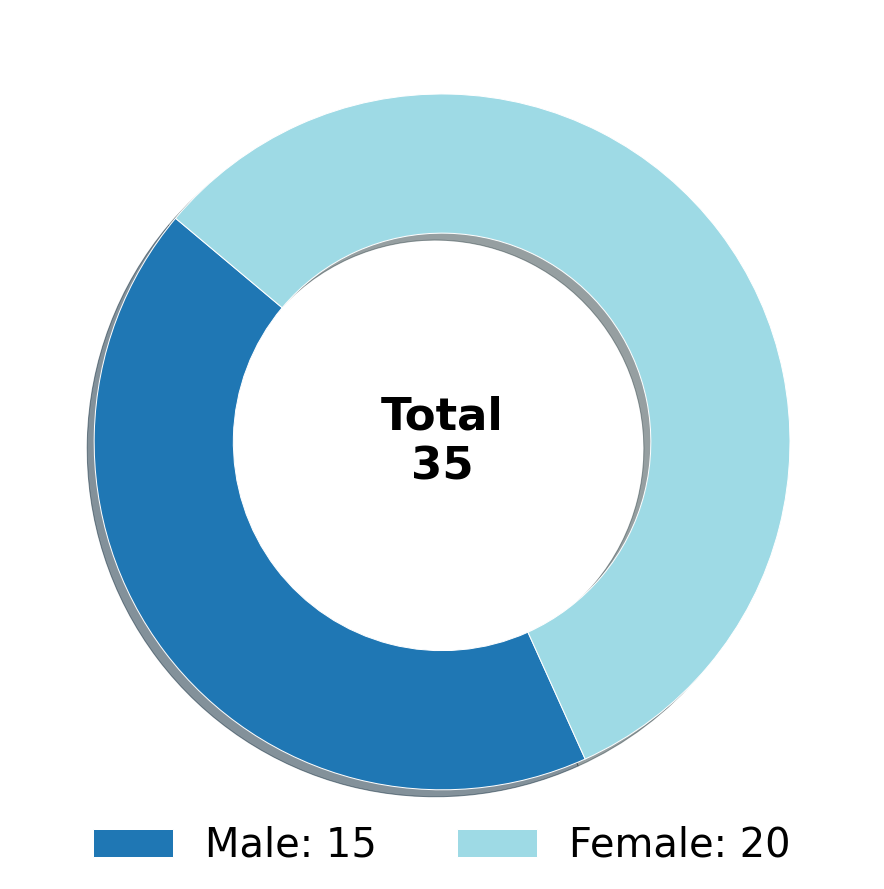}
        \caption{Count for different gender in the test set for annotated data}
        \label{fig:sub5}
    \end{subfigure}
    \hfill
    \begin{subfigure}{0.4\textwidth}
        \includegraphics[width=\linewidth]{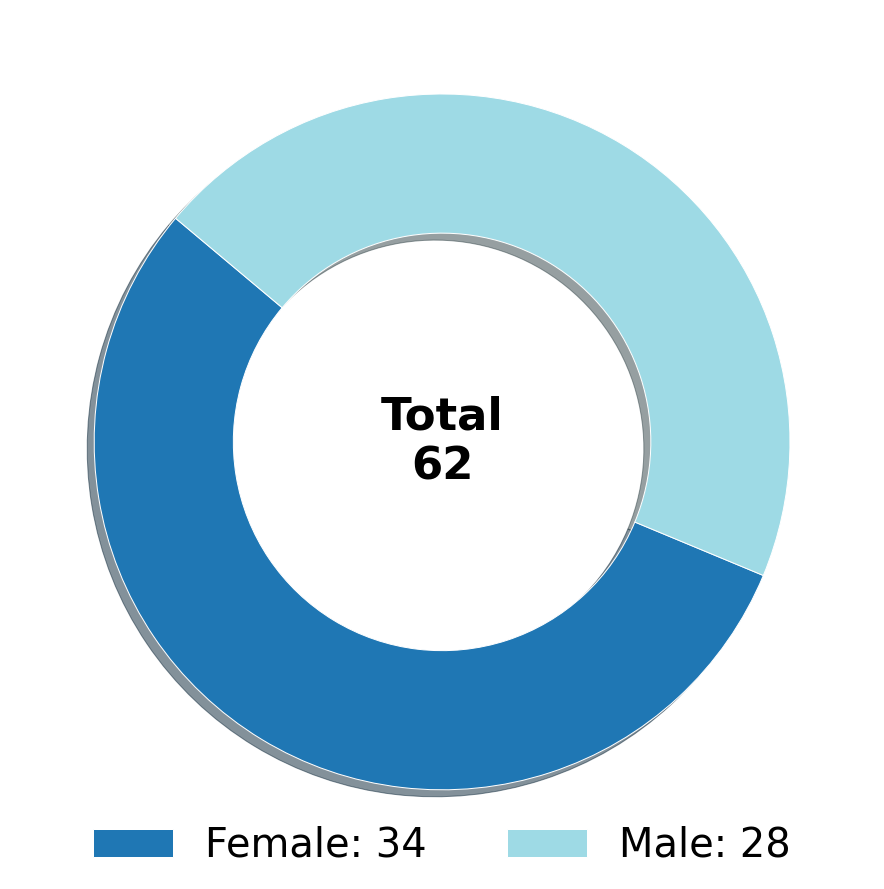}
        \caption{Count for different gender in the test set for annotated data and paired data}
        \label{fig:sub6}
    \end{subfigure}

    \caption{Number of gender in the data set}
    \label{fig:gender}
\end{figure*}

To ensure that our data is sufficiently diverse and that the model can be effectively applied to cases of people of different genders, Figure \ref{fig:gender} demonstrates the distribution of patients' gender in the dataset used in our work. Demographics besides gender are suppressed due to the IRB protocol. 

\subsection{Ablation Study}
\label{sec:ablation}

\begin{table*}[t]
\centering
\begin{tabular}{llll}

\hline
Model                   & Backbone  & Depth-direction strategy     & DSC   \\
\hline
Single-slice model      & MobileSAM & no                           & 84.97 \\
Single-slice model      & Vit-b      & no                           & 84.17 \\
Multi-slice model       & Mobilesam & 3-slice input                & 82.13 \\
Depth-direction Adapter & Mobilesam & add additional depth Adapter & 81.85 \\
Single-slice model      & Vit-b      & our Depth-attention branch   & 86.46 \\
SegmentAnybone          & Mobilesam & our Depth-attention branch   & 86.87 \\
\hline
\end{tabular}
\caption{Ablation study of different components of the model during development.}
\end{table*}
In this section, we present various choices for model component selection during the development phase. Initially, we investigated different network backbones, including Vit-b and MobileSAM. Our findings revealed that the larger image encoder, Vit-b, did not significantly outperform the more compact MobileSAM, regardless of whether we implemented the Depth-attention Branch. Therefore, our paper primarily focuses on the MobileSAM-based models due to their higher computational efficiency.

Additionally, we examined several methods for incorporating 3D information into the segmentation process. One such method is a 2.5D-based model, which uses three adjacent slices as a 3-channel input for the SAM, predicting the central slice. Another approach \citep{wu2023medical}, commonly used in video processing tasks and cited in our references, involved adding a Depth Adapter. However, our experiments indicated that this addition could detract from performance and considerably reduce inference speed due to the transpose and rotation operations within each Attention Block.

\end{multicols}
\end{document}